\documentclass[a4paper,11pt]{article}
\usepackage{jheppub}
\bibstyle{JHEP}

\usepackage[english]{babel} 
\usepackage[utf8]{inputenc}  
\usepackage[T1]{fontenc}
\usepackage{lmodern}
\usepackage{amsmath}         
\usepackage{amsfonts}          
\usepackage{amssymb}
\usepackage{graphicx} 
\usepackage{bbold}
\usepackage{bbm}
\usepackage[colorlinks=true,urlcolor=blue,linkcolor=blue,citecolor=blue,linktocpage=true]{hyperref}

\newcommand{\figref}[1]	{{fig.~\ref{#1}}}
\newcommand{\secref}[1]	{{sec.~\ref{#1}}}
\newcommand{\NN}{{\mathbb{N}}}
\newcommand{\RR}{{\mathbb{R}}}
\newcommand{\ZZ}{{\mathbb{Z}}}
\newcommand{\Tr}{{\mathrm{Tr}}}
\newcommand{\cH}{{\mathcal{H}}}
\newcommand{\cO}{{\mathcal{O}}}
\newcommand{\cL}{{\mathcal{L}}}
\newcommand{\bra}[1] {{\langle #1 |}}
\newcommand{\ket}[1] {{| #1 \rangle}}

\usepackage{tikz}
\usetikzlibrary{calc,decorations.pathmorphing}

\begin{document}

\title{Entwinement as a possible alternative to complexity}
\author{Johanna Erdmenger,}
\author{Marius Gerbershagen}
\affiliation{Institut f{\"u}r Theoretische Physik und Astrophysik \\ and W\"urzburg-Dresden Cluster of Excellence ct.qmat, \\ Julius-Maximilians-Universit{\"a}t W{\"u}rzburg, Am Hubland, 97074 W\"urzburg, Germany}

\abstract{
  Unlike the standard entanglement entropy considered in the holographic context, \emph{entwinement} measures entanglement between degrees of freedom that are not associated to a spatial subregion.
  Entwinement is defined for two-dimensional CFTs with a discrete $\ZZ_N$ gauge symmetry.
  Since the Hilbert space of these CFTs does not factorize into tensor products, even the entanglement entropy associated to a spatial subregion cannot be defined as the von Neumann entropy of a reduced density matrix.
  While earlier works considered embedding the density matrix into a larger, factorizing Hilbert space, we apply a gauge invariant approach by using a density matrix uniquely defined through its relation to the local algebra of observables.
  We furthermore obtain a fully gauge invariant definition of entwinement valid for general CFTs with $\ZZ_N$ gauge symmetry in terms of all observables acting on the degrees of freedom considered.
  Holographically, entwinement is dual to the length of non-minimal geodesics present for conical defects or black holes.
  In this context, we propose a definition of entwinement for thermal states dual to the BTZ black hole.
  Our results show that ``entwinement is enough'' to describe the full bulk geometry for the conical defect and provide strong hints that the same holds true for the BTZ black hole.
  Thus, it provides an alternative to holographic complexity for the theories considered.
}


\maketitle 

\section{Introduction}
The AdS/CFT correspondence \cite{Maldacena:1997re} is one of the most influential developments in the recent history of theoretical physics.
Due to the holographic nature of the correspondence, the encoding of the bulk geometry in the boundary field theory is of particular interest, a question that has received a large amount of attention in the last years.
A strong indication that entanglement must play an important role in this encoding was provided by the seminal work of Ryu and Takayanagi \cite{Ryu:2006bv}.
It connects the entanglement entropy $S(A)$ of a subregion $A$ -- a quantum information quantity in the boundary field theory -- to a geometric quantity in the bulk, namely the area of the minimal codimension two bulk surface $\gamma_A$ anchored at $\partial A$ on the boundary,
\begin{equation}
  S(A) = \frac{\text{Area}(\gamma_A)}{4G_N}.
\end{equation}
Entanglement entropy is defined for a bipartition of the Hilbert space into tensor factors
\begin{equation}
  \cH = \cH_A \otimes \cH_{A^c}
  \label{eq:Hilbert space bipartition}
\end{equation}
as the von Neumann entropy of the reduced density matrix associated to $\cH_A$,
\begin{equation}
  S(A) = -\Tr_{\cH_A}(\rho_A \log \rho_A) \hspace{1cm} \text{where}~\rho_A = \Tr_{\cH_{A^c}}\rho,
  \label{eq:usual definition entanglement entropy}
\end{equation}
while $\gamma_A$ is termed the Ryu-Takayanagi (RT) surface.
The simple and direct connection between the boundary field theory and the bulk geometry provided by the Ryu-Takayanagi proposal has motivated the conjecture that the bulk spacetime is emergent from entanglement, summarized in the slogan ``entanglement builds geometry'' \cite{Swingle:2009bg,VanRaamsdonk:2010pw,Bianchi:2012ev}.

However, there are features of the bulk spacetime that entanglement entropy of spatial subregions $A$ cannot describe.
Among these are regions known as {\it entanglement shadows} at a finite distance from  a  naked singularity, for instance for conical defects. These regions are not probed by the RT surfaces and thus evade a holographic reconstruction of bulk spacetime from the boundary CFT data. Further examples include event horizons of black holes that are not penetrated by any RT surface \cite{Balasubramanian:2014sra,Freivogel:2014lja}. The AdS Schwarzschild and BTZ black holes have an entanglement shadow of thickness of order the AdS scale surrounding the horizon \cite{Hubeny:2013gta}. 
A further example is the growth of wormholes with time in two-sided black hole spacetimes, which entanglement entropy cannot capture \cite{Susskind:2014moa,Susskind:2014rva}.
While the length of the wormhole between the two asymptotic boundaries continues to grow for a long time, the entanglement entropy of a subregion consisting of two parts on both asymptotic boundaries quickly saturates to a constant value \cite{Hartman:2013qma}.

AdS$_3$/CFT$_2$ is a particular simple context where these issues appear, since 2+1 dimensional gravity admits only solutions with constant curvature. Therefore all spacetimes with negative curvature are quotients of pure AdS$_3$.
These quotients are obtained as follows.
In pure AdS$_3$, called the covering space, multiple subregions are identified with each other.
These subregions are referred to as fundamental domains.
For a 2+1 dimensional gravity theory, RT surfaces are geodesics on a constant time slice that end on two points delimiting an interval $A$ on the boundary.
Geodesics on the covering space descend to geodesics on the quotient space.
Since a single point on the quotient space maps to multiple points on the covering space, a single boundary interval $A$ has multiple geodesics attached to it.
The RT surface is the geodesic with minimum length.

\begin{figure}[ht]
  \centering
  \begin{tikzpicture}[scale=0.75]
    \draw (0,0) circle (3cm);
    % tan^2(phi/n)=(n^2*r^2*tan^2(alpha)-L^2)/(n^2*r^2+L^2)  =>  r=L*sqrt((1+tan^2(phi))/(tan^2(alpha)-tan^2(phi)))
    % alpha=0.5=17.19°
    % N=3
    \draw[red] plot[domain=-17.19:17.19,smooth,variable=\phi] ({atan(sqrt((1+tan(\phi)*tan(\phi))/(0.096-tan(\phi)*tan(\phi))))*3/90*sin((\phi+17.19)*3)},{atan(sqrt((1+tan(\phi)*tan(\phi))/(0.096-tan(\phi)*tan(\phi))))*3/90*cos((\phi+17.19)*3)});
    \draw[green] plot[domain=-77.19:77.19,smooth,variable=\phi] ({atan(sqrt((1+tan(\phi)*tan(\phi))/(19.35-tan(\phi)*tan(\phi))))*3/90*sin((\phi+60+17.19)*3)},{atan(sqrt((1+tan(\phi)*tan(\phi))/(19.35-tan(\phi)*tan(\phi))))*3/90*cos((\phi+60+17.19)*3)});
    \draw[blue] plot[domain=-42.81:42.81,smooth,variable=\phi] ({atan(sqrt((1+tan(\phi)*tan(\phi))/(0.8582-tan(\phi)*tan(\phi))))*3/90*sin((\phi-60+17.19)*3)},{atan(sqrt((1+tan(\phi)*tan(\phi))/(0.8582-tan(\phi)*tan(\phi))))*3/90*cos((\phi-60+17.19)*3)});
    \fill[black] (0,0) circle (0.05cm);
    % R_min = 1/tan(alpha_max) = 1/tan(90°/N)
    % arctan(R_min)*radius/90° = Radius des Verschränkungsschattens
    \fill[black,opacity=0.2] (0,0) circle ({{atan(1/tan(90/3))*3/90}});

    % \draw[|-|] (0,3) arc(90:{90-6*17.19}:3) node[midway,right,above] {$B$};
    
    % tan^2(phi)=(r^2*tan^2(alpha)-L^2)/(r^2+L^2)  =>  r=L*sqrt((1+tan^2(phi))/(tan^2(alpha)-tan^2(phi)))
    % alpha=0.5=17.19°
    % N=3
    \begin{scope}[shift={(6,1.5)},scale=0.5]
      \draw (0,0) circle (3cm);
      \foreach \i in {0,1,2}
      { \draw[red] ({{3*sin(\i*120)}},{{3*cos(\i*120)}}) to[out=-90-\i*120,in=-90-\i*120-2*17.19] ({{3*sin(\i*120+2*17.19)}},{{3*cos(\i*120+2*17.19)}});
        \draw[red] ({{3*sin(\i*120)}},{{3*cos(\i*120)}}) arc({{90-\i*120}}:{{90-\i*120-2*17.19}}:3) ({{3*sin(\i*120+2*17.19)}},{{3*cos(\i*120+2*17.19)}});
        \draw ({{3.55*sin(\i*120+17.19)}},{{3.55*cos(\i*120+17.19)}}) node[black] {\footnotesize{$\tilde B^\i_0$}};
        }
      \draw[dashed] (0,0) -- (0,3);
      \draw[dashed] (0,0) -- (2.598,-1.5);
      \draw[dashed] (0,0) -- (-2.598,-1.5);
    \end{scope}
    \begin{scope}[shift={(8,-1.5)},scale=0.5]
      \draw (0,0) circle (3cm);
      \draw[green] (0,3) to[out=-90,in=-90-2*77.19] ({{3*sin(2*77.19)}},{{3*cos(2*77.19)}});
      \draw[green] (0,3) arc(90:90-2*77.19:3) ({{3*sin(2*77.19)}},{{3*cos(2*77.19)}});
      \draw ({{3.55*sin(77.19)}},{{3.55*cos(77.19)}}) node[black] {\footnotesize{$\tilde B^0_1$}}; 
      \draw[dashed] (0,0) -- (0,3);
      \draw[dashed] (0,0) -- (2.598,-1.5);
      \draw[dashed] (0,0) -- (-2.598,-1.5);
    \end{scope}
    \begin{scope}[shift={(10,1.5)},scale=0.5]
      \draw (0,0) circle (3cm);
      \draw[blue] (0,3) to[out=-90,in=-90-2*137.19] ({{3*sin(2*137.19)}},{{3*cos(2*137.19)}});
      \draw[blue] (0,3) arc(90:90-2*137.19:3) ({{3*sin(2*137.19)}},{{3*cos(2*137.19)}});
      \draw ({{3.55*sin(137.19)}},{{3.55*cos(137.19)}}) node[black] {\footnotesize{$\tilde B^0_2$}}; 
      \draw[dashed] (0,0) -- (0,3);
      \draw[dashed] (0,0) -- (2.598,-1.5);
      \draw[dashed] (0,0) -- (-2.598,-1.5);
    \end{scope}
  \end{tikzpicture}
  \caption{LHS: Constant time slice of the conical defect geometry in AdS$_3$ for $N=3$ (i.e. for the quotient group $\mathbbm{Z}_3$), together with minimal (in red) and non-minimal geodesics (in blue and green).
    The entanglement shadow indicated in grey is not penetrated by any minimal geodesic.
    RHS: Some of the corresponding geodesics together with their associated boundary intervals $\tilde B^i_k$ on a constant time slice of the covering space.}
  \label{fig:conical defect}
\end{figure}
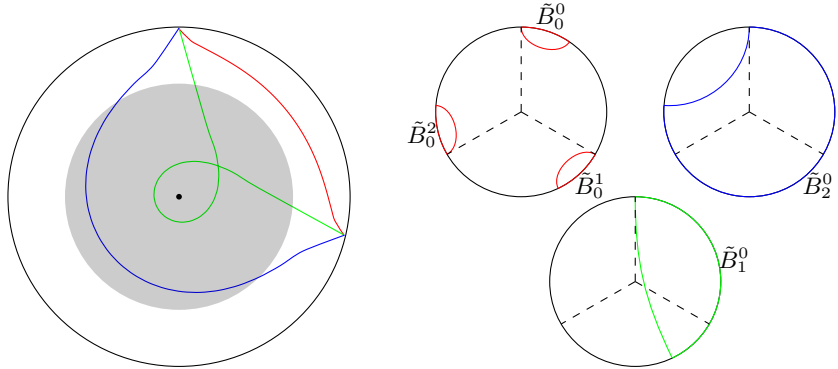

In this work, we mainly focus on conical defects, for which the quotient group is $\ZZ_N$.
In this case, there are $N$ geodesics attached to every boundary interval $A$.
As discussed in \cite{Balasubramanian:2014sra}, the minimal geodesic or RT surface penetrates only to a finite distance into the bulk, leaving an entanglement shadow around the naked singularity at the origin, where there is a region of spacetime not probed by the minimal surfaces that compute spatial entanglement in the dual field theory.
The minimal (RT) geodesics thus do not probe the full spacetime.
On the other hand, as displayed in figure  \figref{fig:conical defect},  the non-minimal geodesics do probe the full spacetime.
This can be most easily seen from the quotienting picture and the fact that geodesics in the covering space correspond to geodesics in the conical defect.
Since the full covering space is probed by geodesics, there are no regions in the conical defect that are not probed by (possibly non-minimal) geodesics.
To completely specify the encoding of the bulk geometry in the boundary field theory from geodesics, it is therefore necessary to obtain a field theory dual to the length of both minimal and non-minimal geodesics.
While it is well established that this dual is the entanglement entropy for minimal geodesics, much less is known about the field theory dual to the length of non-minimal geodesics for which the term \emph{entwinement} has been coined in \cite{Balasubramanian:2014sra}.

Entwinement is a quantum information theoretic quantity in the boundary field theory.
It is defined in \cite{Balasubramanian:2014sra} as the entanglement entropy $E^i_k$ of a boundary interval $\tilde B^i_k$ on the covering space, symmetrized over translations of $\tilde B^i_k$ by elements of the quotient group $\ZZ_N$,
\begin{equation}
  E'_k = \sum_{i \in \ZZ_N}E^i_k.
  \label{eq:definition entwinement Balasubramanian}
\end{equation}
Here, $k$ is a free parameter encoding the size of $\tilde B^i_k$.
To be precise, $\tilde B^i_k$ is given as the union of $k$ fundamental domains with an interval $B$ contained in a single fundamental domain (see \figref{fig:conical defect}).
While the degrees of freedom in consideration are localized in $\tilde B^i_k$ on the covering space, they are not localized in a single subregion in the quotient space.
Thus, entwinement is proposed as a measure for entanglement between non-spatially organized degrees of freedom \cite{Balasubramanian:2014sra}, i.e.~degrees of freedom that are not associated to a spatial subregion on the boundary.

There is however an important subtlety that a definition of any field theory quantity needs to address, namely all quantities must be gauge invariant.
In the conical defect case, the boundary field theory naturally aquires a $\ZZ_N$ gauge symmetry \cite{Balasubramanian:2014sra}, for which the factorization \eqref{eq:Hilbert space bipartition} of the Hilbert space into tensor factors does not hold and reduced density matrices for spatial subregions are not well defined a priori.
Hence, even finding a field theory dual to the length of a minimal geodesic first requires determining a suitable generalization of the reduced density matrix and of entanglement entropy.
Furthermore, turning back to entwinement we note that while the definition \eqref{eq:definition entwinement Balasubramanian} correctly reproduces the length of a non-minimal geodesic, it requires introducing unphysical non-$\ZZ_N$ invariant states that live in the field theory on the boundary of the covering space.
We  also note that while we have defined entwinement for holographic theories, it is in principle a well-defined quantity for any field theory with  $\ZZ_N$ gauge symmetry.

In this paper, we develop fully gauge invariant techniques to characterize entanglement for CFTs with discrete $\ZZ_N$ gauge symmetries.
These techniques are based on the sets of local operators acting on the degrees of freedom considered.
They apply to both entanglement entropy for spatial subregions and to entwinement and do not rely on introducing unphysical gauge invariant states.
For the entanglement entropy associated to a spatial subregion, the set of local operators forms an algebra.
This enables us to apply well-known techniques (see e.g.~\cite{Harlow:2016vwg,OhyaPetz1993} for a review) to define a unique reduced density matrix, whose von Neumann entropy reproduces the RT formula.
For entwinement, the set of operators does not form an algebra.
Instead, it forms only a linear subspace that is closed under addition but not multiplication, as was already observed in \cite{Balasubramanian:2018ajb}.
We develop a new technique to define an entropy associated to this linear subspace.
We define entwinement as the minimum of the entropy of a probability distribution obtained from projective measurements using operators from the linear subspace.
We show that this new definition gives the same results as those of \cite{Balasubramanian:2014sra} described above.
Therefore, we obtain a gauge invariant field theory dual to the length of both minimal and non-minimal geodesics in the conical defect.

Moreover, we propose a definition of entwinement for thermal states of the CFT dual to the BTZ black hole.
The BTZ black hole is dual to a thermal state, i.e.~to an Euclidean CFT on the torus of spatial size $2\pi$.
We propose that entwinement is given as the entanglement entropy of an interval on a ``large'' torus of spatial size $2\pi M$, where $M \in \NN$.
We show that its bulk dual probes not only the entanglement shadows around the horizon, but also the growth of the wormhole for the two-sided black hole.
Under some assumptions about the microscopic nature of the field theory description of black holes, the proposed definition of entwinement is a well-defined quantity in the field theory on the small torus with spatial size $2\pi$ and in accord with the definition using a linear subspace.

This result is particularly interesting in light of the holographic complexity conjectures of \cite{Susskind:2014moa,Susskind:2014rva}.
Complexity is a quantity from quantum information which measures how difficult it is to prepare a target state from a given reference state by applying certain unitary gates.
Several possible holographic duals to complexity have been proposed in \cite{Susskind:2014rva,Stanford:2014jda,Brown:2015bva,Brown:2015lvg}.
It is, however, still unclear which complexity measures in the field theory reproduce these proposals, despite much work in this direction (see e.g.~\cite{Chapman:2017rqy,Jefferson:2017sdb,Khan:2018rzm,Hackl:2018ptj,Chapman:2018hou,Goto:2018iay} and references therein).
It has been conjectured that complexity can explain the features of the bulk geometry not captured by the entanglement entropy of spatial subregions, in particular the growth of the wormhole in two-sided black hole geometries \cite{Susskind:2014moa,Susskind:2014rva}.
Entwinement as defined in this paper probes the same features of the bulk geometry that complexity is conjectured to explain.
In particular, these features include the wormhole growth with time in the two-sided black hole case.
Therefore, at least for the geometries considered -- conical defects in AdS$_3$ and the BTZ black hole -- it provides an alternative to complexity.

Other work on entwinement includes \cite{Balasubramanian:2016xho,Balasubramanian:2018ajb}.
In \cite{Balasubramanian:2016xho}, entwinement was calculated for symmetric product orbifolds using a replica trick and agreement with the length of non-minimal bulk geodesics was found.
The authors of \cite{Balasubramanian:2018ajb} pointed out similarities between entwinement and entanglement for indistinguishable particles.
They also suggested that entwinement is naturally associated to a linear subspace of observables, a proposal that we verify in our calculations.

Our paper is organized as follows:
We begin in \secref{sec:CFT review} with a review of  CFTs with discrete gauge symmetry.
In \secref{sec:entanglement entropy} we apply the algebraic approach to defining entanglement entropy to these CFTs.
We show equivalence of this approach both with a naive calculation using unphysical non-gauge invariant states and for holographic CFTs with the Ryu-Takayanagi proposal.
In \secref{sec:entwinement}, we turn to the study of entwinement.
We introduce a gauge invariant definition of entwinement based on a linear subspace of observables.
We show that this method is in agreement both with earlier calculations of \cite{Balasubramanian:2014sra,Balasubramanian:2016xho} using non-gauge invariant states and with the length of non-minimal geodesics in the bulk.
Sec.~\ref{sec:BTZ} contains a proposal for entwinement in the thermal state of the CFT dual to the BTZ black hole.
We specify the assumptions underlying this proposal and show that it implies that entwinement probes not only the entanglement shadows around the horizon, but also the growth of the wormhole for the two-sided black hole.
In \secref{sec:S_n Orbifolds}, we briefly comment on $S_n$ orbifolds arising from the D1-D5 system.
Finally in \secref{sec:conclusions}, we present our conclusions.

\section{Conformal field theories with discrete gauge symmetry}
\label{sec:CFT review}
This section serves as an introduction to the conformal field theories whose entanglement structure will be determined below.
In particular, we will introduce tensor product theories and permutation orbifolds and present their Hilbert space structure.

We begin by considering the conical defect as a quotient of pure AdS$_3$.
This pure AdS$_3$, the covering space, is partitioned into $N$ fundamental domains which are identified under the action of the quotient group.
This implies that the boundary splits into $N$ fundamental domains as well.
On each fundamental domain lives a conformal field theory with central charge $\tilde c$, which we will call the \emph{seed CFT}.
Taking $N$ non-interacting copies of this seed CFT without changing the boundary conditions obeyed by the fields of the CFTs gives a \emph{tensor product CFT} denoted as CFT$^N$.

Coming back from the holographic setting to general conformal field theories, we consider as an example the seed CFT given by $\tilde c$ free bosons $X^{m}(t,\phi)$ on a circle with action
\begin{equation}
  S_\text{CFT}[X] = \sum_{m=0}^{\tilde c-1} \int dt \int_0^{2\pi}d\phi\, \partial X^{m}(t,\phi)\bar\partial X^{m}(t,\phi),
  \label{eq:action seed CFT}
\end{equation}
and periodic boundary conditions $X^m(t,\phi + 2\pi) = X^m(t,\phi)$.
The tensor product theory is a CFT of $c = N\tilde c$ free bosons $X^{m,n}(t,\phi)$ with action
\begin{equation}
  S_{\text{CFT}^N}[X] = \sum_{m=0}^{\tilde c-1}\sum_{n=0}^{N-1} \int dt \int_0^{2\pi}d\phi\, \partial X^{m,n}(t,\phi)\bar\partial X^{m,n}(t,\phi),
  \label{eq:action tensor product CFT}
\end{equation}
and boundary conditions $X^{m,n}(t,\phi + 2\pi) = X^{m,n}(t,\phi)$.

In general, the Hilbert space of the tensor product theory is a $N$-fold tensor product of the Hilbert space of the seed CFT,
\begin{equation}
  \cH_{\text{CFT}^N} = (\cH_\text{CFT})^{\otimes N}.
\end{equation} 
To implement the quotient, we have to demand that our theory be invariant under $\ZZ_N$ permutations of the copies.
It is well known how to construct such $\ZZ_N$ invariant CFTs, which go under the name of \emph{permutation orbifolds}\footnote{Such orbifolds can be constructed for general permutation groups. We comment on entanglement in the $S_n$ case in \secref{sec:S_n Orbifolds}.} (for an introduction to CFT orbifolds in general see for example \cite{Ginsparg:1988ui,Dixon:1986qv}).
We will be mainly interested in the Hilbert space $\cH_\text{orb}$ of the orbifold theory, which splits up into twisted sectors $\cH_{\text{orb},k}$ labeled by $k \in \ZZ_N$,
\begin{equation}
  \cH_\text{orb} = \bigoplus_{k=0}^{N-1}\cH_\text{orb,k} \, ,
\end{equation}
in which the fields obey the boundary conditions
\begin{equation}
  X^{m,n}(t,\phi + 2\pi) = X^{m,n+k}(t,\phi),
  \label{eq:boundary conditions twisted sector}
\end{equation}
where $n+k$ is understood to be modulo $N$.
As is  important for the consideration of entanglement in these CFTs, each of the $\cH_{\text{orb},k}$ sectors contains only $\ZZ_N$ symmetric states.
This means that in each $\cH_{\text{orb},k}$ sector we apply the projection operator
\begin{equation}
  P = \frac{1}{N} \sum_{k \in \ZZ_N} g^k,
\end{equation}
where $g^k$ are the $\ZZ_N$ generators acting on the fields as
\begin{equation}
  g^k X^{m,n} g^{-k} = X^{m,n+k}.
\end{equation}
Only $\ZZ_N$ symmetric states with $g$ eigenvalue 1 survive the projection.
The conical defect geometry is identified with the ground state of the $k=1$ twisted sector.
The central charge $c$ of the orbifold theory is unchanged as  compared to the covering theory.

A further theory will prove useful later on, although not directly describing the actual physical systems we are interested in.
This theory is obtained from the tensor product theory by changing the boundary conditions of the fields to be periodic up to permutations
\begin{equation}
  X^{m,n}(t,\phi + 2\pi) = X^{m,n+1}(t,\phi).
  \label{eq:boundary conditions covering CFT}
\end{equation}
The theory in which the fields obey \eqref{eq:boundary conditions covering CFT} is called the \emph{covering CFT}.
It has the same action as the tensor product theory, which in our example using boundary conditions \eqref{eq:boundary conditions covering CFT} can be equivalently written as the theory of $c/N = \tilde c$ free bosons $\tilde X^m(t,\phi)$ on a $N$ times larger spatial circle,
\begin{equation}
  S_\text{covering CFT}[X] = \sum_{m=0}^{\tilde c-1} \int dt \int_0^{2\pi N}d\phi\, \partial \tilde X^m(t,\phi) \bar\partial \tilde X^m(t,\phi).
  \label{eq:action covering CFT}
\end{equation}
While this theory can describe aspects of states contained in the $k=1$ twisted sector of the orbifold theory, it is not equivalent to the orbifold theory restricted to the $k=1$ twisted sector.
In particular, its Hilbert space contains non-$\ZZ_N$ invariant states.
The $k=1$ twisted sector of the orbifold is only equivalent to the covering theory when the $\ZZ_N$ symmetry of the action \eqref{eq:action covering CFT} is promoted to a gauge symmetry.

\section{Entanglement entropy}
\label{sec:entanglement entropy}
Entanglement entropy is commonly defined as \eqref{eq:usual definition entanglement entropy} using the bipartition \eqref{eq:Hilbert space bipartition} of the Hilbert space into tensor factors.
A direct application of this definition is not possible in our case, since the Hilbert space of the orbifold theory does not decompose into tensor factors.
After showing this for the case of a $\ZZ_2$ symmetric Hilbert space, we explain two methods for dealing with this problem and prove that they give equivalent results.
Finally, we generalize to the $\ZZ_N$ case.
It is interesting to note that similar problems occur in lattice gauge theories and analogous methods to the ones presented here have been employed to define entanglement entropy in these systems\footnote{We would like to thank Mari-Carmen Bañuls for pointing this out to us.} (see for example \cite{Buividovich:2008gq,Donnelly:2011hn,Casini:2013rba,Radicevic:2014kqa,Aoki:2015bsa,Ghosh:2015iwa,Soni:2015yga}).

The Hilbert space of the CFT$^2$ tensor product theory is given by
\begin{equation}
  \cH = \left\{\, \ket{X,Y} \,\right\},
\end{equation}
where we have denoted field eigenstates as $\ket{X,Y}$ such that
\begin{equation}
  X^{m,1}(\phi)\ket{X,Y} = X^m(\phi)\ket{X,Y}, ~~ X^{m,2}(\phi)\ket{X,Y} = Y^m(\phi)\ket{X,Y}.
\end{equation}
The $\ZZ_2$ symmetry generators are $g^0=\mathbb{1}$ and
$g$ with action
\begin{equation}
  g\ket{X,Y} = \ket{Y,X}.
\end{equation}
$\cH$ decomposes in $\ZZ_2$ symmetric and antisymmetric states
obtained by applying the orthogonal projection operators
$P^\pm=\frac{1}{2}(\mathbb{1} \pm g)$,
\begin{equation}
  P^\pm\ket{XY} = \frac{1}{2}(\ket{X,Y} \pm \ket{Y,X}),
\end{equation}
with $P^\pm P^\mp = 0$ and $(P^\pm)^2 = P^\pm$. Hence
\begin{equation}
  \cH = \cH^+ \oplus \cH^-
\end{equation}
with
\begin{equation}
  \cH^\pm = P^\pm\cH = \left\{~ (\ket{X,Y} \pm \ket{Y,X})/2 ~\right\}.
\end{equation}
The orbifold twisted sectors are given by $\cH^+$ with appropriate
boundary conditions \eqref{eq:boundary conditions twisted sector}. We
will restrict to states in a single sector and drop the $k$ index
labeling the different sectors. This does not introduce any problems,
since the states we are interested in lie in a single twisted sector.
Moreover, states in different twisted sectors are orthogonal to each
other, thus in traces of the density matrix over the Hilbert space
$\cH_\text{orb}$, only the twisted sector containing the state we are
interested in contributes. 

The Hilbert spaces of the tensor product and covering theories decompose into tensor factors associated to a subregion $A$,
\begin{equation}
  \cH = \cH_A \otimes \cH_{A^c}, 
\end{equation}
with $A^c$ the complement of $A$. 
We denote the field eigenstates spanning $\cH_A$ by $\ket{X_A,Y_A}$
and analogously for $\cH_{A^c}$, such that a state in $\cH$ can
alternatively be written as
$\ket{X,Y} = \ket{X_AX_{A^c},Y_AY_{A^c}}$.\footnote{This
  notation just means that
  \begin{equation}
    X^{m,1}(\phi)\ket{X_AX_{A^c},Y_AY_{A^c}} = \left\{
      \begin{aligned}
        X_A(\phi)~&,~~\phi \in A\\
        X_{A^c}(\phi)~&,~~\phi \in A^c
      \end{aligned}
    \right.
  \end{equation}
  and analogously for $X^{m,2}$.}  The $\cH^+$ Hilbert space of the
orbifold, on the other hand, does not decompose into tensor factors. This
can be seen by considering the Hilbert space of states that are
$\ZZ_2$ (anti-)symmetrized only in the subregion $A$ or $A^c$,
\begin{equation}
  \cH_A^\pm = \left\{~ (\ket{X_A,Y_A} \pm \ket{Y_A,X_A})/2 ~\right\}, ~~
  \cH_{A^c}^\pm = \left\{~ (\ket{X_{A^c},Y_{A^c}} \pm \ket{Y_{A^c},X_{A^c}})/2 ~\right\}.
\end{equation}
Tensoring a state of $\cH_A^+$ with a state of $\cH_{A^c}^+$ gives
\begin{equation}
  \begin{aligned}
    \frac{1}{4}(\ket{X_A,Y_A} + \ket{Y_A,X_A}) \otimes (\ket{X'_{A^c},Y'_{A^c}} + \ket{Y'_{A^c},X'_{A^c}})\\
    = \frac{P^+}{2}(\ket{X_AX'_{A^c},Y_AY'_{A^c}} + \ket{X_AY'_{A^c},Y_AX'_{A^c}})
  \end{aligned}
  \label{eq:H+ decomposition 1}
\end{equation}
which is manifestly in $\cH^+$. But tensoring a state of $\cH_A^-$
with one of $\cH_{A^c}^-$ also gives a state of $\cH^+$,
\begin{equation}
  \begin{aligned}
    \frac{1}{4}(\ket{X_A,Y_A} - \ket{Y_A,X_A}) \otimes (\ket{X'_{A^c},Y'_{A^c}} - \ket{Y'_{A^c},X'_{A^c}})\\
    = \frac{P^+}{2}(\ket{X_AX'_{A^c},Y_AY'_{A^c}} - \ket{X_AY'_{A^c},Y_AX'_{A^c}}).
  \end{aligned}
  \label{eq:H+ decomposition 2}
\end{equation}
From the explicit form of the r.h.s.~of \eqref{eq:H+ decomposition 1},
\eqref{eq:H+ decomposition 2}, it is obvious that pairs of these
states span all of $\cH^+$, hence $\cH^+$ is given as a direct sum
\begin{equation}
  \cH^+ = (\cH_A^+ \otimes \cH_{A^c}^+) \oplus (\cH_A^- \otimes \cH_{A^c}^-).
\end{equation}
Similarly,
\begin{equation}
  \cH^- = (\cH_A^+ \otimes \cH_{A^c}^-) \oplus (\cH_A^- \otimes \cH_{A^c}^+).
\end{equation}
Therefore, the physical Hilbert space $\cH^+$ of $\ZZ_2$ symmetric
states decomposes only into a direct sum of tensor factors.

\subsection{Embedding the state into a larger Hilbert space}
\label{sec:embedding EE}
A simple way of dealing with the problem of a non-factorizing Hilbert space is to embed states into the enlarged Hilbert space $\cH$, which factorizes along spatial degrees of freedom. Explicitely, we use the density matrix $\rho \oplus 0$ in $\cH = \cH^+ \oplus \cH^-$ to define a reduced density matrix and an entanglement entropy $S(A)$ by
\begin{equation}
  S(A) = \Tr_{\cH_A}(\rho_A \log \rho_A)   \, ,\hspace{1cm} \text{where}~\rho_A = \Tr_{\cH_{A^c}}(\rho \oplus 0).
  \label{eq:density matrix EE}
\end{equation}
In practice, this means that we compute $S(A)$ as the entanglement entropy of the union of all copies $\tilde A^i$ of $A$ in the covering theory,
\begin{equation}
  S(A) = \tilde S(\bigcup_{i=0}^{N-1} \tilde A^i),
  \label{eq:EE as union of EE in covering theory}
\end{equation}
where the $\tilde S$ notation is a reminder that this entanglement entropy is computed in the covering theory.
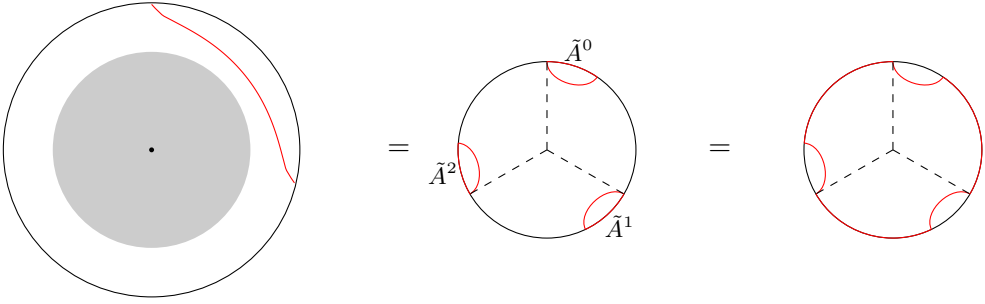
\begin{figure}
  \centering
  \begin{tikzpicture}[scale=0.65]
    \draw (0,0) circle (3cm);
    % tan^2(phi/n)=(n^2*r^2*tan^2(alpha)-L^2)/(n^2*r^2+L^2)  =>  r=L*sqrt((1+tan^2(phi))/(tan^2(alpha)-tan^2(phi)))
    % alpha=0.5=17.19°
    % N=3
    \draw[red] plot[domain=-17.19:17.19,smooth,variable=\phi] ({atan(sqrt((1+tan(\phi)*tan(\phi))/(0.096-tan(\phi)*tan(\phi))))*3/90*sin((\phi+17.19)*3)},{atan(sqrt((1+tan(\phi)*tan(\phi))/(0.096-tan(\phi)*tan(\phi))))*3/90*cos((\phi+17.19)*3)});
    \fill[black] (0,0) circle (0.05cm);
    % R_min = 1/tan(alpha_max) = 1/tan(90°/N)
    % arctan(R_min)*radius/90° = Radius des Verschränkungsschattens
    \fill[black,opacity=0.2] (0,0) circle ({{atan(1/tan(90/3))*3/90}});

    % \draw[|-|] (0,3) arc(90:{90-6*17.19}:3) node[midway,right,above] {$B$};
    
    % tan^2(phi)=(r^2*tan^2(alpha)-L^2)/(r^2+L^2)  =>  r=L*sqrt((1+tan^2(phi))/(tan^2(alpha)-tan^2(phi)))
    % alpha=0.5=17.19°
    % N=3
    \draw (5,0) node {$=$};
    \begin{scope}[shift={(8,0)},scale=0.6]
      \draw (0,0) circle (3cm);
      \foreach \i in {0,1,2}
      { \draw[red] ({{3*sin(\i*120)}},{{3*cos(\i*120)}}) to[out=-90-\i*120,in=-90-\i*120-2*17.19] ({{3*sin(\i*120+2*17.19)}},{{3*cos(\i*120+2*17.19)}});
        \draw[red] ({{3*sin(\i*120)}},{{3*cos(\i*120)}}) arc({{90-\i*120}}:{{90-\i*120-2*17.19}}:3) ({{3*sin(\i*120+2*17.19)}},{{3*cos(\i*120+2*17.19)}});
        \draw ({{3.6*sin(\i*120+17.19)}},{{3.55*cos(\i*120+17.19)}}) node[black] {\footnotesize{$\tilde A^\i$}};
      }
      \draw[dashed] (0,0) -- (0,3);
      \draw[dashed] (0,0) -- (2.598,-1.5);
      \draw[dashed] (0,0) -- (-2.598,-1.5);
    \end{scope}
    \draw (11.5,0) node {$=$};
    \begin{scope}[shift={(15,0)},scale=0.6]
      \draw (0,0) circle (3cm);
      \foreach \i in {0,1,2}
      { \draw[red] ({{3*sin(\i*120)}},{{3*cos(\i*120)}}) to[out=-90-\i*120,in=-90-\i*120-2*17.19] ({{3*sin(\i*120+2*17.19)}},{{3*cos(\i*120+2*17.19)}});
        \draw[red] ({{3*sin(\i*120+2*17.19)}},{{3*cos(\i*120+2*17.19)}}) arc({{90-\i*120-2*17.19}}:{{90-\i*120-120}}:3) ({{3*sin(\i*120-120)}},{{3*cos(\i*120-120)}});
        %\draw ({{3.75*sin(\i*120+60+17.19)}},{{3.55*cos(\i*120+60+17.19)}}) node[black] {\footnotesize{$(\tilde A^\i)^c$}};
      }
      \draw[dashed] (0,0) -- (0,3);
      \draw[dashed] (0,0) -- (2.598,-1.5);
      \draw[dashed] (0,0) -- (-2.598,-1.5);
    \end{scope}
  \end{tikzpicture}
  \caption{The entanglement entropy of a subregion $A$ in the conical defect is given by the entanglement entropy of the union of all copies of $A$ in the covering space or equivalently its complement.
    For a small interval, the RT surface is the union of the RT surfaces of the copies of $A$, while for a large interval it is the union of the RT surfaces of the complement.
    This exchange of dominance is the reason for the emergence of the entanglement shadow, depicted in grey on the left.}
  \label{fig:entanglement entropy conical defect}
\end{figure}
This procedure has been used to define entanglement entropy for the conical defect in \cite{Balasubramanian:2014sra} and for holographic CFTs reproduces the RT proposal (see \figref{fig:entanglement entropy conical defect}).
However,  it requires introducing unphysical states that are traced over to obtain the reduced density matrix.
As we now show, there exists a different approach which bypasses these problems and in fact leads to the same results.

\subsection{Algebraic entanglement entropy}
We begin with a review of the algebraic approach to defining entanglement entropy \cite{Harlow:2016vwg,OhyaPetz1993} for general systems, before applying it to the $\ZZ_2$ and $\ZZ_N$ orbifold theories introduced in \secref{sec:CFT review}.
The basic idea of this approach is to consider the unique density matrix $\rho_M \in M$ associated to a von Neumann algebra $M$ by the requirement that it gives the same expectation values as the global density matrix for all operators from $M$,
\begin{equation}
  \Tr_{\cH^0}(\rho \cO) = \Tr_{\cH^0}(\rho_M \cO) ~~ \forall \cO \in M.
  \label{eq:density matrix algebra}
\end{equation}
Here we have denoted the gauge invariant Hilbert space as $\cH^0$ (the special case of a $\ZZ_2$ gauge symmetry for which $\cH^0 \equiv \cH^+$ is described in the subsection \ref{sec:ZZ_2 EE}).
Then, the entanglement entropy associated to $M$ is given by the von Neumann entropy for $\rho_M$,
\begin{equation}
  S(M) = \hat\Tr_{\cH^0}(\rho_M \log \rho_M),
  \label{eq:EE algebra}
\end{equation}
where $\hat\Tr$ indicates that the trace is normalized differently than the $\Tr$ functional, essentially due to the fact that we trace over the whole Hilbert space instead of only a tensor factor as in \eqref{eq:usual definition entanglement entropy}.

For an introduction to algebraic entanglement entropy for finite-dimensional Hilbert spaces see \cite{Harlow:2016vwg}, whose conventions we will use below.
The statements that we will make in the following are valid only for finite-dimensional Hilbert spaces (type I von Neumann algebras).
This does not introduce any additional problems, since the entanglement entropy is strictly infinite for type II and type III von Neumann algebras on infinite dimensional Hilbert spaces \cite{Witten:2018lha,OhyaPetz1993}, thus a regularization is needed in any case.
A general, rigorous treatment of algebraic entanglement entropy can be found in \cite{OhyaPetz1993}.

The algebras of local operators that we consider are von Neumann algebras, i.e.~sets of operators that are closed under Hermitean conjugation, addition and multiplication and contain the identity.
The set of operators $Z_M \subset M$ that commute with all elements of a von Neumann algebra $M$ is called the center of $M$.
A von Neumann algebra with trivial center is called a factor.
The importance of factor algebras is that they define an associated Hilbert space decomposition into tensor factors,
\begin{equation}
  \cH = \cH_A \otimes \cH_{A^c}.
\end{equation}
The elements of a factor algebra $M_A$ of operators localized in $A$ act trivially on $\cH_{A^c}$, i.e.~$M_A$ is given as
\begin{equation}
  M_A = \cL(\cH_A) \otimes \mathbb{1},
\end{equation}
where $\cL(\cH_A)$ is the set of linear operators acting on the Hilbert space $\cH_A$.
Hence, if $M_A$ has trivial center the algebraic definition reduces to the usual one and the density matrix associated to $M_A$ is given the reduced density matrix on $\cH_A$ times the identity on $\cH_{A^c}$,
\begin{equation}
  \rho_{M_A} = \rho_A \otimes \frac{\mathbb{1}_{\cH_{A^c}}}{|\cH_{A^c}|},
\end{equation}
with $\rho_A = \Tr_{\cH_{A^c}}\rho$.
The $1/|\cH_{A^c}|$ factor is necessary for the correct normalization $\Tr\rho_{M_A}=1$.

For von Neumann algebras $M_A$ with non-trivial center acting on the Hilbert space of gauge invariant states $\cH^0$, $Z_M$ is in general spanned by a set of projection operators $Q^\alpha$.
Since these are mutually commuting, they can be simultaneously diagonalized.
Therefore, the projection operators can be taken to be mutually orthogonal, $Q^\alpha Q^\beta = Q^\alpha$ for $\alpha = \beta$ and $Q^\alpha Q^\beta = 0$ otherwise.
These operators project onto factors, i.e.~$Q^\alpha M_A Q^\alpha$ is a factor on
$Q^\alpha \cH^0$ while $Q^\alpha M_A Q^\beta = 0$ for $\alpha \neq \beta$.
This implies a factorization of the Hilbert space
\begin{equation}
  \cH^0 = \bigoplus_\alpha (\cH^\alpha_A \otimes \cH^\alpha_{A^c}),
  \label{eq:HH+ Hilbert space decomposition}
\end{equation}
while $M_A$ decomposes as
\begin{equation}
  M_A = \bigoplus_\alpha (\cL(\cH^\alpha_A) \otimes \mathbb{1}_{\cH^\alpha_{A^c}}) \, .
  \label{eq:algebra of observables acting on HH+}
\end{equation}
This decomposition is just the explicit form of the statement that $M_A$ acts locally in $A$ and maps $\cH^0$ into itself.
The reduced density matrix $\rho_{M_A}$ associated to $M_A$ is given by
\begin{equation}
  \rho_{M_A} = \bigoplus_\alpha \left( \rho^\alpha_A \otimes \frac{\mathbb{1}_{\cH^\alpha_{A^c}}}{|\cH^\alpha_{A^c}|} \right) \, ,
  \label{eq:rho M A in algebraic approach}
\end{equation}
where
\begin{equation}
  \rho^\alpha_A = \Tr_{\cH^\alpha_{A^c}}(Q^\alpha \rho \, Q^\alpha).
  \label{eq:rho alpha A in algebraic approach}
\end{equation}
Remember that $Q^\alpha$ is the projection onto $\cH^\alpha_A \otimes \cH^\alpha_{A^c}$, i.e.~$Q^\alpha \cH^0 = \cH^\alpha_A \otimes \cH^\alpha_{A^c}$.
Therefore, the entanglement entropy associated to $M_A$ is given by
\begin{equation}
  S(M_A) = -\hat\Tr_{\cH^0}(\rho_{M_A} \log \rho_{M_A}) = -\sum_\alpha \Tr_{\cH^\alpha_A}(\rho^\alpha_A \log \rho^\alpha_A).
  \label{eq:algebraic EE}
\end{equation}
The effect of the modified trace is to cancel out the $1/|\cH^\alpha_{A^c}|$ factors, which were needed to ensure the correct normalization of $\rho_{M_A}$.
We can already see that this modification is necessary by considering the case that $M_A$ is a factor.
If we had not cancelled the $1/|\cH_{A^c}|$ factor, the definition \eqref{eq:EE algebra} would not agree with \eqref{eq:usual definition entanglement entropy}.

\subsubsection{$\ZZ_2$ case}
\label{sec:ZZ_2 EE}
In the case we are considering, $\alpha \in \{+,-\}$ and the projections are given by
\begin{equation}
  Q^\pm = \frac{1}{2}(\mathbb{1} \pm g_A) = \frac{1}{2}(\mathbb{1} \pm g_{A^c}),
\end{equation}
where $g_A,g_{A^c}$ are $\ZZ_2$ generators acting only in the subregion $A,A^c$\, ,
\begin{equation}
  g_A\ket{X_AX_{A^c},Y_AY_{A^c}} = \ket{Y_AX_{A^c},X_AY_{A^c}}, ~~
  g_{A^c}\ket{X_AX_{A^c},Y_AY_{A^c}} = \ket{X_AY_{A^c},Y_AX_{A^c}}.
\end{equation}
We expand $\rho$ in field eigenstates
\footnote{
  $\rho(X,Y,X',Y')$ is a functional of the functions $X,Y,X',Y'$.
  The integration measures $dXdY$ and $dX'dY'$ are assumed to be appropriately normalized.},
\begin{equation}
  \rho = \int dXdY \int dX'dY' \rho(X,Y,X',Y') \ket{X,Y} \bra{X',Y'},
  \label{eq:rho expanded in field eigenstates}
\end{equation}
where $\rho(X,Y,X',Y')$ obeys
\begin{equation}
  \rho(X,Y,X',Y') = \rho(Y,X,X',Y') = \rho(X,Y,Y',X') = \rho(Y,X,Y',X')
  \label{eq:rho ZZ_2 symmetry}
\end{equation}
due to the fact that $\rho$ is a density matrix for states in $\cH^+$ and thus $\rho = g\rho = \rho g$.
We obtain for $\rho_A^\pm$ that
\begin{equation}
  \begin{aligned}
    \rho_A^\pm = \Tr_{\cH^\pm_{A^c}}(Q^\pm \rho Q^\pm) = \Tr_{\cH^\pm_{A^c}}(Q^\pm \rho) \\
    = \frac{1}{8} \int dXdY \int dX'_AdY'_A &(\rho(X,Y,X'_AX_{A^c},Y'_AY_{A^c}) \pm \rho(X,Y,X'_AY_{A^c},Y'_AX_{A^c}))\\
    &\times(\ket{X_A,Y_A} \pm \ket{Y_A,X_A})(\bra{X'_A,Y'_A} \pm \bra{Y'_A,X'_A})
\end{aligned}
\end{equation}
and therefore
\begin{equation}
  \begin{aligned}
    \rho_A &= \rho_A^+ \oplus \rho_A^-\\
    &= \frac{1}{4} \int dXdY \int dX'_AdY'_A\,\ket{X_A,Y_A} \bra{X'_A,Y'_A}\,
    \begin{aligned}[t]
      (&\rho(X_AX_{A^c},Y_AY_{A^c},X'_AX_{A^c},Y'_AY_{A^c})\\
      +&\rho(Y_AX_{A^c},X_AY_{A^c},Y'_AX_{A^c},X'_AY_{A^c})\\
      +&\rho(X_AX_{A^c},Y_AY_{A^c},Y'_AY_{A^c},X'_AX_{A^c})\\
      +&\rho(Y_AX_{A^c},X_AY_{A^c},X'_AY_{A^c},Y'_AX_{A^c}))
    \end{aligned}\\
    &= \int dXdY \int dX'_AdY'_A \,\rho(X_AX_{A^c},Y_AY_{A^c},X'_AX_{A^c},Y'_AY_{A^c})\ket{X_A,Y_A} \bra{X'_A,Y'_A} \, ,
  \end{aligned}
  \label{eq:rho_A final result}
\end{equation}
where in the last line we have used \eqref{eq:rho ZZ_2 symmetry}.
Eq.~\eqref{eq:algebraic EE} together with $\cH_A = \cH_A^+ \oplus \cH_A^-$ then implies that the von Neumann entropy of $\rho_A$ is equal to the entanglement entropy associated to $M_A$ in the algebraic approach.
Furthermore, from the last line of \eqref{eq:rho_A final result}, it is clear that $\rho_A$ is nothing but the reduced density matrix obtained in section \ref{sec:embedding EE} by first embedding $\rho$ in $\cH = \cH^+ \oplus \cH^-$ as $\rho \oplus 0$ and then taking the partial trace over $\cH_{A^c}$ \footnote{Compare for example to the definition of a partial trace of a density matrix $\rho = \rho_{ij,kl}\ket{ij}\bra{kl}$ of a two qubit system, $\rho_A = \rho_{ik,jk}\ket{i}\bra{j}$. The last line of \eqref{eq:rho_A final result} has exactly the same structure, where the sum over $k$ is the equivalent of the integral over $X_{A^c},Y_{A^c}$.}.
Therefore, we have shown that for the $\ZZ_2$ case, the algebraic approach and the embedding procedure of the last subsection give the same results.

\subsubsection{$\ZZ_N$ case}
The generalization of the above procedure to the $\ZZ_N$ case is straightforward. 
As above, the Hilbert space $\cH$ of the covering theory decomposes into a direct sum $\cH = \bigoplus_{n=0}^{N-1} \cH^n$ of Hilbert spaces $\cH^n$ with fixed $g$ eigenvalues of $e^{-2\pi in/N}$.
The $\cH^n$ are obtained from $\cH$ by applying the projection operators
\begin{equation}
  P^n = \frac{1}{N} \sum_{j=0}^{N-1}e^{2\pi inj/N}g^j \, .
\end{equation}
As for the $P^\pm$ above, the $P^n$ are pairwise orthogonal and square to themselves, $P^nP^m = P^n\delta^{nm}$.
The Hilbert space of the $\ZZ_N$ orbifold is given as a direct sum of $\cH^0$ Hilbert spaces with appropriate twisted boundary conditions.
Each $\cH^n$ factor decomposes into a direct sum of tensor products
\begin{equation}
  \cH^n = \bigoplus_{m=0}^{N-1} \cH_A^m \otimes \cH_{A^c}^{n-m} \, ,
  \label{eq:direct sum ZN Hilbert space}
\end{equation}
where the $\cH_A^n$ and $\cH_{A^c}^n$ factors are obtained from $\cH_A$ and $\cH_{A^c}$ by applying the projection operators
\begin{equation}
  Q_{A,A^c}^n = \frac{1}{N}\sum_{j=0}^{N-1} e^{2\pi inj/N}g_{A,A^c}^j.
\end{equation}
Applying \eqref{eq:rho alpha A in algebraic approach} together with \eqref{eq:direct sum ZN Hilbert space} for $n=0$, we obtain
\begin{equation}
  \rho_A^n = \Tr_{\cH_{A^c}^{N-n}}(Q_A^n \rho Q_A^n) \, .
\end{equation}
For $\rho \in \cH^0$ we again obtain
\begin{equation}
  \bigoplus_{n=0}^{N-1} \rho_A^n = \rho_A \, ,
\end{equation}
where $\rho_A$ is given by embedding $\rho$ into $\cH$ and then tracing out $\cH_{A^c}$.
Hence also for the $\ZZ_N$ case, we obtain equality between the entanglement entropy in the orbifold and covering theory.

This justifies the calculation procedure for the entanglement entropy in the conical defect as the entanglement entropy of the union of all copies of $A$ in the covering space from \eqref{eq:EE as union of EE in covering theory}.
It also implies that in the specific example that we studied -- even though a decomposition of the Hilbert space into tensor factors $\cH_A\otimes\cH_{A^c}$ does not exist -- the RT proposal remains applicable if we identify the entanglement entropy $S(A)$ with the algebraic entanglement entropy \eqref{eq:algebraic EE}.

\section{Entwinement}
\label{sec:entwinement}
Sec.~\ref{sec:entanglement entropy} aimed at answering the question how entangled the degrees of freedom in a subregion $A$ are with those in its complement $A^c$.
However, this is not the only question that one can ask about the entanglement structure of a density matrix.
Other interesting questions are about the entanglement structure of a non-spatially organized subset of the degrees of freedom, e.g.~the fields of one of the $N$ copies of the seed CFT, with the remainder of the system\footnote{This subset and its complement is an example of a bipartition of the target space, i.e.~the space of field values of the theory. Entanglement with respect to target space bipartitions has recently been studied in \cite{Mazenc:2019ety}.}.
In particular, for entwinement we are concerned with the entanglement between degrees of freedom localized in different subregions for each copy of the seed CFT.

From the definition \eqref{eq:definition entwinement Balasubramanian} of \cite{Balasubramanian:2014sra}, entwinement is given by a sum of entanglement entropies $E^i_k$ in the covering space.
All $E^i_k$ in the sum give the same contribution, therefore it is sufficient to obtain just one of them for calculating entwinement.
In fact as we will see later on, a gauge invariant definition of entwinement can not distinguish the $E^i_k$ for different $i$.
All $E^i_k$ are obtained from the same linear subspace of gauge invariant operators, thus the physical quantity that the $E^i_k$ represent is the same for all $i$.
Therefore, we will drop the $i$ index in the following and define entwinement to be the quantity
\begin{equation}
  E_k = E^i_k ~~~~ \text{with}~ i = 0,...,N-1 ~\text{arbitrary}.
  \label{eq:definition entwinement}
\end{equation}
As a consequence, our definition \eqref{eq:definition entwinement} differs by a factor of $1/N$ from the definition \eqref{eq:definition entwinement Balasubramanian} of \cite{Balasubramanian:2014sra}.

As before, the Hilbert space does not decompose into tensor factors for the degrees of freedom whose entanglement entropy $E^i_k$ represents.
To solve this problem, we adopt the same methods as for entanglement entropy in \secref{sec:entanglement entropy}.
It turns out that the first method described above of embedding the state into a larger, factorizing Hilbert space works fine, while the second method of using the algebraic entanglement entropy does not work in the entwinement case.
As already observed in \cite{Balasubramanian:2018ajb}, the set of observables associated to entwinement only forms a linear subspace instead of an algebra.
Going beyond the work of \cite{Balasubramanian:2018ajb}, we develop a gauge invariant definition of entwinement from measurements with operators of the aforementioned linear subspace.

\subsection{Embedding the state into a larger Hilbert space}
As for ordinary entanglement entropy, we embed the density matrix into
the Hilbert space $\cH = \bigoplus_{n=0}^{N-1}\cH^n$ of all states
regardless of their gauge invariance properties. Field eigenstates of
$\cH$ are denoted as $\ket{X^0...X^{N-1}}$. Since we do not perform any
symmetrization on these states, the corresponding Hilbert space
factorizes,
\begin{equation}
  \cH = \cH_{\tilde B_k} \otimes \cH_{\tilde B^c_k},
\end{equation}
where $\tilde B_k = \tilde B^i_k$ for any $i=0,...,N-1$. A state that
is unentangled with respect to this factorization is given by
\begin{equation}
  \ket{X^0...X^{N-1}} = \ket{X^0...X^{k-1}X^k_B} \otimes \ket{X^k_{B^c}X^{k+1}...X^{N-1}},
  \label{eq:state factorization}
\end{equation}
where as before $\ket{X^n_B}$ denotes a field eigenstate for the
$X^{m,n}$ fields defined in the subregion $B$. Due to the
factorization property the reduced density matrix
\begin{equation}
  \rho_{\tilde B_k} = \Tr_{\cH_{\tilde B^c_k}}(\rho \oplus 0 \oplus ... \oplus 0)
  \label{eq:reduced density matrix entwinement}
\end{equation}
is well-defined and thus $E_k$ is obtained as
\begin{equation}
  E_k = \Tr_{\cH_{\tilde B_k}}(\rho_{\tilde B_k} \log \rho_{\tilde B_k}) \, .
  \label{eq:definition entwinement covering space}
\end{equation}
The important difference between $\rho_{\tilde B_k}$ and a reduced
density matrix $\rho_A$ for ordinary entanglement entropy is that in
\eqref{eq:reduced density matrix entwinement} we have traced out
different fields over different subregions, while in \eqref{eq:density
  matrix EE} all fields were traced out over the same subregion. Of
course, this calculation is unsatisfactory for the same reason as that
for ordinary entanglement entropy: In taking the partial trace, we
trace out unphysical non-gauge invariant states.

\subsection{Entwinement from linear subspace of observables}
In contrast to the entanglement entropy studied in the last subsection, for entwinement there is no associated algebra of observables in the orbifold theory.
Starting from the simple $\ZZ_2$ case, we first determine that the set of operators acting on the degrees of freedom associated to entwinement forms only a linear subspace that is closed under addition, but not multiplication.
Then we show that nevertheless there is an entropy associated to this linear subspace which is equal to the entwinement $E_k$.

\subsubsection{$\ZZ_2$ case}
We focus again on the simple example with $N=2$.
In the covering theory, we can associate an algebra $M_{\tilde B_k}$ of observables to entwinement, which contains operators acting in the full spatial circle on $k$ fields and in the subregion $B$ on one field.
Explicitly,
\begin{equation}
  M_{\tilde B_k} = \cL(\cH_{\tilde B_k}) \otimes \mathbb{1},
\end{equation}
where $\cL(\cH_{\tilde B_k})$ is the set of linear operators acting on
$\cH_{\tilde B_k}$. For the example of $k=1$, $\cL(\cH_{\tilde B_1})$
is given by operators of the form
\begin{equation}
  \cO_{\tilde B_1} = \int dXdY_B \int dX'dY'_B \,\cO(X,Y_B,X',Y'_B) \ket{X,Y_B} \bra{X',Y'_B}.
\end{equation}
Since the elements of the $M_{\tilde B_k}$ algebra are operators in
the covering theory, the algebra doesn't map $\cH^+$ into
itself. To obtain operators that map $\cH^+$ into itself, we have to
project onto operators that are block diagonal with respect to the
factorization $\cH = \cH^+ \oplus \cH^-$:\footnote{The algebra $M_A$
  of operators in $A$ acting on $\cH^+$ introduced in the last
  subsection is obtained in the same way from the algebra of operators
  in $A$ acting on $\cH$.}
\begin{equation}
  M_B = P^+M_{\tilde B_k}P^+ + P^-M_{\tilde B_k}P^- = \frac{1}{2}\left(\cL(\cH_{\tilde B_k}) \otimes \mathbb{1} + g(\cL(\cH_{\tilde B_k}) \otimes \mathbb{1})g\right).
  \label{eq:lin subspace for entwinement}
\end{equation}
The $+$ in \eqref{eq:lin subspace for entwinement} means that every
element of $M_B$ is a sum of an element of $M_{\tilde B_k}$ with the
corresponding $\ZZ_2$ translated element of $gM_{\tilde B_k}g$, i.e.
\begin{equation}
  M_B = \left\{~ \frac{1}{2}\left(\cO_{\tilde B_k} \otimes \mathbb{1} + g(\cO_{\tilde B_k} \otimes \mathbb{1})g\right) ~\right\}.
\end{equation}
The basic problem now is that $M_B$ does not form an algebra, since it
is not closed under multiplication. As already observed in
\cite{Balasubramanian:2018ajb}, $M_B$ forms a linear subspace
that is closed only under addition and Hermitean
conjugation. Therefore, the method which we used in the previous
subsection to define the entanglement entropy no longer applies. A
direct generalization of eq.~\eqref{eq:density matrix algebra} to
linear subspaces is also not possible due to the structure of the
Hilbert space. Unlike the case of ordinary entanglement entropy, the
$\ZZ_2$ symmetric Hilbert space $\cH^+$ does not possess a
decomposition as a tensor product or a direct sum of tensor products
in terms of which $\rho_{\tilde B_k}$ is a density matrix for some
subset of states in $\cH^+$.

However, $\rho_{\tilde B_k}$ still is associated to the linear
subspace $M_B$ in the sense that $\rho_{\tilde B_k}$ is the part of
the full density matrix $\rho$ that can be measured using operators
from $M_B$. Before showing this, we would like to briefly remind the
reader of some basic concepts from quantum information theory. Every
density matrix $\rho$ can written in terms of a spectral decomposition
\begin{equation}
  \rho = \sum_i p_i \ket{\psi_i} \bra{\psi_i},
\end{equation}
where $p_i \in [0,1]$ is the probability that the state of the quantum
system is $\ket{\psi_i}$. The $\ket{\psi_i}$ form a complete
orthonormal basis for the Hilbert space of the system. Performing a
projective measurement with a Hermitean operator
$\cO = \sum_j m_j \ket{\chi_j} \bra{\chi_j}$ yields the state
\begin{equation}
  \rho_\cO = \sum_{i,j} p_i |\langle\chi_j|\psi_i\rangle|^2 \ket{\chi_j} \bra{\chi_j} = \sum_{j} p_{\cO,j} \ket{\chi_j} \bra{\chi_j},
\end{equation}
where $p_{\cO,j}$ is the probability of measuring the eigenvalue $m_j$ of $\cO$.
The von Neumann entropy of $\rho_\cO$ is in general
larger than the von Neumann entropy of $\rho$ \cite{NielsenChuang},
\begin{equation}
  S(\rho_\cO) = S(\mathbf{p}_\cO) = -\sum_jp_{\cO,j}\log p_{\cO,j} \geq S(\rho) = S(\mathbf{p}) =  -\sum_ip_i\log p_i.
  \label{eq:measurement increases entropy}
\end{equation} 
Equality is obtained when the eigenbases of $\cO$ and $\rho$ agree,
i.e.~when $\ket{\psi_i} = \ket{\chi_i}$. In this case, we have full
knowledge of the density matrix $\rho$. Hence, the von Neumann entropy
of $\rho$ can be obtained by looking for the infimum of $S(\rho_\cO)$
over the set of all Hermitean operators or equivalently over the set
of bases $\ket{\chi_j}$,
\begin{equation}
  S(\rho) = \inf\{ S(\mathbf{p}_\cO) ~|~ \text{$\{\ket{\chi_j}\}$ is a basis for the Hilbert space of $\rho$} \}.
  \label{eq:entropy from infimum over measurements}
\end{equation}

\medskip
The same procedure can be applied if one has access only to the linear
subspace $M_B$ of Hermitean operators. $M_B$ is spanned by a set of
basis operators of the form
\begin{equation}
  \cO_j = \frac{1}{2} \left(\ket{\chi_j}\bra{\chi_j} \otimes \mathbb{1} + g(\ket{\chi_j}\bra{\chi_j} \otimes \mathbb{1})g\right),
\end{equation}
where the $\ket{\chi_j}$ form a complete orthonormal basis for the covering
Hilbert space factor $\cH_{\tilde B_k}$. Again, measurements
using the operators $\cO_j$ yield a probability distribution given by
\begin{equation}
  p_{\cO,j} = \Tr_{\cH^+}(\rho \cO_j)
  \label{eq:probability distribution entwinement}
\end{equation}
and a corresponding entropy $S(\mathbf{p}_\cO) = -\sum_jp_{\cO,j}\log p_{\cO,j}$.
We define entwinement to be the infimum of $S(\mathbf{p}_\cO)$ over all bases
$\{\cO_j\}$,
\begin{equation}
  E_k = S(M_B) = \inf\{ S(\mathbf{p}_\cO) ~|~ \text{$\{\cO_j\}$ is a basis for $M_B$} \}.
  \label{eq:definition entwinement linear subspace}
\end{equation}
Due to \eqref{eq:entropy from infimum over measurements}, to show that
the definition \eqref{eq:definition entwinement linear subspace} agrees with \eqref{eq:definition entwinement covering
  space}, we only have to show that the probability distribution
\eqref{eq:probability distribution entwinement} agrees with the
probability distribution $\tilde p_{\ket{\chi_j}\bra{\chi_j}}$ for the
basis $\ket{\chi_j}$ with respect to $\rho_{\tilde B_k}$. This follows
directly from the fact that $g \rho g=\rho$ and $\rho\cH^- = 0$,
\begin{equation}
  p_{\cO,j} = \Tr_{\cH^+}\left( \rho\, \cO_j \right) = \Tr_\cH(\rho\, \ket{\chi_j}\bra{\chi_j} \otimes \mathbb{1}) = \Tr_{\cH_{\tilde B_k}}(\rho_{\tilde B_k} \ket{\chi_j}\bra{\chi_j}) = \tilde p_{\ket{\chi_j}\bra{\chi_j}}.
\end{equation}
Thus we see that entwinement is obtained as the minimal entropy
of the probability distribution of a measurement with an operator from
$M_B$. Therefore, entwinement is a measure for the amount of
information that can be obtained about the density matrix $\rho$ from
measurements with operators in the linear subspace $M_B$.

We further note that this definition of entwinement reduces to the
entanglement entropy from the algebraic approach if the linear
subspace closes into an algebra, at least for algebras on 
finite-dimensional Hilbert spaces considered in the last section. This
follows directly from the fact that the algebras $M_A$ we consider are
linear subspaces as well, therefore we can apply the same techniques
as above. For each $\cH^\alpha_A$ factor of the decomposition
\eqref{eq:HH+ Hilbert space decomposition}, we choose a basis of
states, which yields a basis for the Hermitean operators of $M_A$. Due
to \eqref{eq:density matrix algebra}, the probability distributions
for a measurement of these operators with respect to $\rho_{M_A}$ and
$\rho$ agree. Hence, the minimal entropy of this probability
distribution is equal to the von Neumann entropy of $\rho_{M_A}$ given
by eq.~\eqref{eq:EE algebra}. Therefore, the definition
\eqref{eq:definition entwinement linear subspace} of an entropy
associated to a linear subspace of operators is a generalization of
the algebraic approach, which itself is a generalization of the usual
definition of entanglement entropy using a partial trace.

\subsubsection{$\ZZ_N$ case} 
The generalization of the above method to the $\ZZ_N$ case follows immediately.
The projection of the elements of the algebra $M_{\tilde B_k}$ acting on $\cH$ to operators that are block-diagonal for the $\cH = \bigoplus_n\cH^n$ decomposition gives the linear subspace
\begin{equation}
  M_B = \sum_{n=0}^{N-1} P^n M_{\tilde B_k} P^n = \frac{1}{N} \sum_{j=0}^{N-1} g^j M_{\tilde B_k} g^{N-j}.
  \label{eq:ZZ_N subspace}
\end{equation}
As before, the Hilbert space $\cH^0$ does not decompose into tensor
products for which a reduced density matrix associated to $M_B$ could
be defined. Due to $g\rho = \rho g = \rho$, expectation values
w.r.t.~$\rho$ for elements
\begin{equation}
  \cO_j = \frac{1}{N} \sum_{j=0}^{N-1} g^j (\ket{\chi_j}\bra{\chi_j}\otimes\mathbb{1}) g^{N-j}
\end{equation}
of $M_B$ are equal to expectation values for $\ket{\chi_j}\bra{\chi_j}$ w.r.t.~$\rho_{\tilde B_k}$,
\begin{equation}
  p_{\cO_j} = \Tr_{\cH^0}(\rho \cO_j) = \Tr_\cH(\rho \ket{\chi_j}\bra{\chi_j}\otimes\mathbb{1}) = \Tr_{\cH_{\tilde B_k}}(\rho_{\tilde B_k} \ket{\chi_j}\bra{\chi_j}) = \tilde p_{\ket{\chi_j}\bra{\chi_j}}.
\end{equation}
Here $\ket{\chi_j}$ is a basis for $\cH_{\tilde B_k}$. Thus we see
that also in the $\ZZ_N$ case, $E_k$ is given as the minimal entropy
of the probability distribution of measurements with operators from
$M_B$.

Since $M_B$ is independent of the choice of subregion $\tilde B^i_k$
in the covering space on which $\ket{\chi_j}\bra{\chi_j}$ acts, $E_k$
is independent of this choice as well. Alternatively, we see that from
the perspective of measurements with gauge invariant operators, all
$E^i_k$ factors from \eqref{eq:definition entwinement Balasubramanian}
are associated to the same linear subspace and thus represent the same
physical quantity. Hence, the length of a non-minimal geodesic is in
fact not directly equal to the entropy associated to $M_B$ divided by
$4G_N$, but only up to a prefactor of $1/N$.

So far we have calculated entwinement associated to the degrees of
freedom (up to the $\ZZ_N$ symmetry) $X^0,...,X^{k-1},X_B^{k}$,
i.e.~$k$ fields on the full space and one field on the subregion
$B$. However, the definition of entwinement given above is
sufficiently general to allow for different subsets of the total
degrees of freedom $X^0,...,X^{N-1}$. For example, consider the subset
consisting of the union of $N/M$ subsets
$X^{Mi},...,X^{Mi+k-1},X_B^{Mi+k}$ labeled by $i \in {0,...,N/M-1}$,
where $M \in \NN$ divides $N$ and $k < M$. Holographically, the
corresponding quantity is the union of $N/M$ geodesics on the covering
space. From the results of the last section, it is clear that this is
equivalent to a single minimal geodesic in a conical defect geometry
that arises from quotienting the covering space by a $\ZZ_{N/M}$
group. The corresponding central charge is $\tilde c N/M = c/M$. Of
course, the geodesic will be non-minimal in the original $\ZZ_N$
quotient conical defect if $k \geq 1$\footnote{Note however that now
  the possibility of a phase transition arises, just
  as for the ordinary entanglement entropy.}. Thus we see that using
another covering space which is obtained by unrolling the spatial
coordinate $N/M$ times instead of $N$ times will still give us the
length of a non-minimal geodesic, altough with a different prefactor
$c/M$ instead of $c/N$ for $E_k$. For $k=0$ and $M=1$, the geodesic
becomes minimal and we recover the ordinary entanglement entropy.

\subsubsection{Relation to existing results} 
Generalizations of entanglement similar to the one derived above have been considered before in \cite{2005IJTP...44.2127B,2003PhRvA..68c2308B}.
There, entanglement has been defined with respect to a convex cone $C$ of states, i.e.~a set of operators representing density matrices that is closed under taking convex linear combinations $\lambda x_1 + (1-\lambda)x_2$, where $\lambda \in [0,1]$.
Pure states are unit trace elements of $C$ that cannot be written as convex combinations of other elements of $C$.
Furthermore, a second cone $D$ is introduced together with a map $\pi$ from $C$ to $D$.
Then, a pure state $x \in C$ is termed {\it generalized unentangled} relative to $D$ if $\pi(x)$ is pure as well.
The authors of \cite{2005IJTP...44.2127B,2003PhRvA..68c2308B} also introduce an entanglement measure on states $y \in D$ as
\begin{equation}
  S(y) = \inf\{ S(\mathbf{p}) ~|~ y = \sum_i p_iy_i ~\text{with $y_i$ pure} \},
  \label{eq:entanglement measure convex cones}
\end{equation}
where $S(\mathbf{p})$ is a Schur concave entropy measure on the probability distribution $\{p_i\}$, e.g.~the Shannon entropy $S(\mathbf{p}) = -\sum_ip_i\log p_i$.
This definition is equivalent to the ordinary entanglement if $D$ is the set of operators acting on a tensor factor of the Hilbert space that $C$ acts on and if $\pi$ is the partial trace\footnote{The proof of this statements works by showing that the probability distribution of any decomposition of $y$ is a transformation of the probability distribution of the spectral decomposition of $y$ by a doubly stochastic matrix, which increases the value of all Schur concave functions.}.
It is also equivalent to entwinement if we identify $C$ with the set of operators acting on $\cH^0$ and $D$ with $M_B$.
In this case, the map $\pi$ takes an element of $C$ to an element of $D$ by embedding it in the enlarged Hilbert space $\cH$, taking the partial trace over $\cH_{\tilde B_k^c}$ and projecting onto a $\ZZ_N$ invariant operator as in \eqref{eq:ZZ_N subspace}.
Using the convex cone formalism, one can in fact show in general that a restriction to measurements using a subspace of observables implies the existence of a pair of cones $C$ and $D$ for which the generalized entanglement described above can be defined \cite{2005IJTP...44.2127B}.

The entanglement measure \eqref{eq:entanglement measure convex cones} is extended to mixed states in \cite{2005IJTP...44.2127B,2003PhRvA..68c2308B} using the convex hull construction familiar from entanglement of formation, i.e.~the entanglement measure for mixed states is given by a second infimum over convex decompositions of the full state in $C$.
We do not extend the entwinement definition in the same way, since we do not aim at separating quantum correlations due to entanglement from classical correlations due to a mixed global density matrix.
Rather, we are interested in quantifying the total amount of correlations measured using operators from a linear subspace.
The reason for that stems from the intuition from ordinary entanglement entropy:
The Ryu-Takayanagi formula applies for both mixed and pure states, although ordinary entanglement entropy quantifies the total amount of correlations in a given subsystem and not those from entanglement alone.

\section{Application to the BTZ black hole}
\label{sec:BTZ}
As for the conical defect, the BTZ black hole \cite{Banados:1992wn} is a simple quotient of pure AdS$_3$.
Therefore we expect to be able to apply the same techniques as in the previous section.
In the following we propose a definition of entwinement for thermal states and show that for simple examples of boundary subregions it reproduces the length of non-minimal geodesics in the dual BTZ black hole spacetime.
These geodesics probe important features of the bulk geometry, in particular not only the entanglement shadow but also the growth of the wormhole for the two-sided black hole.
Therefore entwinement offers a possible alternative to complexity to describe these bulk geometry features from boundary data.
The definition we give rests on some assumptions about the microscopic nature of the dual CFT state, which we will detail but not attempt to verify.
We only focus on the non-rotating black hole.

\subsubsection*{The covering theory}

Since black holes are dual to thermal states, the corresponding CFT is Euclidean with compactified time, i.e.~the CFT lives on the torus.
The BTZ black hole arises from a quotient of pure AdS$_3$ with $\ZZ$ quotient group.
The starting point is pure AdS$_3$ in coordinates for which the metric is
\begin{equation}
  ds^2=-\left(\frac{r^2}{L^2}-m\right)dt^2+\left(\frac{r^2}{L^2}-m\right)^{-1}dr^2+r^2d\phi^2,
  \label{eq:BTZ metric}
\end{equation}
where $m$ is the mass of the black hole and $\phi \in \RR$.
The BTZ black hole is obtained by identifying $\phi \sim \phi + 2\pi$.
Prior to this identification, the boundary theory is in the Rindler vacuum with respect to the $t,\phi$ coordinates \cite{Maldacena:1998bw}.
Thus we expect in the BTZ case the covering theory to be a CFT on a line (spatial coordinate $\phi \in \RR$) in a thermal state.
Therefore, we propose the covering theory to be the Euclidean CFT on a cylinder with periodic time direction.
From the $\ZZ$ quotient group, the covering theory acquires a $\ZZ$ gauge symmetry.
The cylinder with periodic time can be thought of as being formed by an infinite number of tori cut open at a fixed value of the spatial coordinate and glued together.
Then, the $\ZZ$ gauge symmetry acts by permuting these tori with each other.

For the massless black hole, which can be obtained either as the $m \to 0$ limit of the BTZ geometry or the $N \to \infty$ limit of the conical defect, the above definition for the covering field theory is in accord with the one for the conical defect.
The $m \to 0$ limit implies that $\beta$, the radius of the circle in the time direction, goes to infinity.
Thus we are left with a CFT on a line with non-compact time and $\ZZ$ symmetry permuting strips with infinite temporal and finite spatial extent with each other, which is precisely the $N \to \infty$ limit of the covering theory for the conical defect.

There is however a problem in defining the covering theory since this requires unwrapping the spatial direction an infinite number of times.
Therefore, for fixed $c$ the central charge $\tilde c = c/N$ of the covering theory is zero, leading to an ill-defined theory.
But as we have seen in the last section, it is also possible to use a covering space which covers the original geometry $M < N$ times, where $N$ is the number of copies in the full covering space which is infinite in the BTZ case.
On the CFT side, this implies that instead of unwrapping the torus an infinite number of times to a cylinder, we only take a finite number $M \ll c$ of torus copies.
To summarize, for the covering theory to be well-defined, we take it to be a CFT on a torus which is $M$ times larger in the space direction than the one of the quotient theory.
Since $\tilde c = c/M \gg 1$, this \emph{finite covering CFT} is still in the large $c$ limit.

For this definition to be in accord with the one from \secref{sec:entwinement}, however, we have to make some assumptions about the microscopic nature of the field theory description of the BTZ black hole.
The BTZ spacetime is not dual to a single field theory state, but to an ensemble of microstates.
The above definition is only well-defined if (on average) each microstate is in a twisted sector of the theory, such that the set of fields of the theory split into $M$ subsets which are continuously connected by the boundary conditions.
Furthermore, computing entwinement in each microstate and averaging over the result must reproduce the definition given above.
For the massless BTZ black hole, these assumptions have been verified to hold  for $M \lesssim \sqrt c$ in the CFT of the D1/D5-brane system at the orbifold point \cite{Balasubramanian:2016xho}.
In the massive black hole case, we leave the question for which values of $M$ this assumption is justified, or whether it is justified at all, for future work.
Instead we will explore the implications of our entwinement proposal for the dual bulk description in terms of geodesics.

For simplicity, we restrict our calculation of entwinement to entangling intervals in the finite covering theory that are small compared to the size of the spatial direction of the torus.
This means that we work in the $k \ll M$ limit, where $k$ is the parameter of \eqref{eq:definition entwinement}.
In this limit, the results for the entanglement entropy in the finite covering CFT will match those of the entanglement entropy of the same interval in a CFT with the same finite $\tilde c$ on the cylinder with periodic time.
Therefore, we can equivalently take our finite covering theory to be the theory on the cylinder with periodic time and finite $\tilde c$.
Then we simply calculate the entanglement entropy of an interval in this finite covering theory.

\subsubsection*{One-sided black hole}

Concentrate first on the one-sided BTZ black hole.
As described above, for $k \ll M$ to obtain entwinement we only need to calculate the entanglement entropy in the finite covering theory, i.e.~a CFT on the cylinder with periodic time and finite $\tilde c$.
Thus, we have to compute the entanglement entropy of a single interval of length $L=|B|+2\pi k$ on the cylinder with periodic time, where $|B|<2\pi$ is the length of the interval $B$ and $k \in \NN$.
The result for this entanglement entropy was derived in \cite{Calabrese:2004eu} and is given by, up to a regularization constant
\begin{equation}
  E_k = \frac{\tilde c}{3} \log\left( \frac{\beta}{2\pi} \sinh\left(
  \frac{\pi L}{\beta} \right) \right),
  \label{eq:EE cylinder}
\end{equation}
where $\beta$ is the inverse temperature.
The interpretation of this result is as follows.
  Eq.~\eqref{eq:EE cylinder} is equal to the entanglement entropy on the cylinder with periodic time for any $k \in \NN$.
  Eq.~\eqref{eq:EE cylinder} can also be identified with entwinement, but as explained above, this identification holds only for $k \ll M \ll c$.
  This can be checked by referring to the entanglement entropy on the finite covering space (the large torus with period $2\pi M$).
  The entanglement entropy in this case has been computed in \cite{Barrella:2013wja} and as expected agrees with \eqref{eq:EE cylinder} as long as $k \ll M$.
  Finally, eq.~\eqref{eq:EE cylinder} is equal to $1/4G_NM$ times the length of a geodesic associated to a boundary interval of length $|B|$ and winding $k$ times around the black hole horizon.
  The equality between geodesic lengths and eq.~\eqref{eq:EE cylinder} holds for all $k \in \NN$.
  However, for entwinement $k \ll M \ll c$, hence there is an upper bound on the winding number of those geodesics that are dual to entwinement.
For high winding numbers, the geodesics come very close to the black hole horizon.
Therefore, the breakdown between geodesic lengths and entwinement in this limit might be a signal that bulk quantum corrections must be taken into account to describe the region near the black hole horizon.

\subsubsection*{Two-sided black hole}
  We now turn to the case of the two-sided BTZ black hole, which is dual to the thermofield double state \cite{Maldacena:2001kr},
  \begin{equation}
    \ket{\Omega} = \frac{1}{\sqrt{Z}} \sum_{n,m} e^{-\beta E_n/2} \ket{E_n}_1 \ket{E_n}_2\, , ~~~~ \text{with} ~ \ket{E_n} ~ \text{an energy eigenstate}.
  \end{equation}
  The same arguments as for the one-sided case lead to the finite covering theory being the CFT on a line with central charge $\tilde c$ in the thermofield double state.
  The reason is that correlation functions in the thermofield double state (including correlation functions of twist operators that yield entanglement entropy via the replica trick) are calculated by correlation functions of the Euclidean theory in a background with periodic time $\tau$.
  The difference to the thermal case is that operators on the first boundary are inserted at $\tau=0$, while operators on the second boundary are inserted at $\tau=i\beta/2$. 
  Therefore, we can apply the same mapping to the covering theory as for the one-sided case described in detail above.
  The only difference to be taken into account is that in correlation functions, operators on the second asymptotic boundary are inserted at $\tau=i\beta/2$.
  For simplicity, we will also restrict to the same small interval limit $k \ll M$ as above.

Since the entanglement entropy in the thermofield double state for an interval living on only one boundary reduces to the value in the thermal state, interesting entangling intervals are those that consist of two components on different asymptotic boundaries.
The entanglement entropy for a CFT on a line in the thermofield double state was calculated in \cite{Hartman:2013qma}.
For an interval consisting of two subintervals of length $L = |B| + 2\pi k$ at the same position on both asymptotic boundaries, the authors of \cite{Hartman:2013qma} obtain a growth of the entanglement entropy for a time of order $t \sim L/2$ followed by a saturation to a constant value,
\begin{equation}
  E_k = \left\{
  \begin{aligned}
    \frac{2\tilde c}{3} \log\left( \frac{\beta}{2\pi} 2\sinh\left(\frac{\pi L}{\beta} \right)\right)~&,~~t \gg L/2 \, ,\\
    \frac{2\tilde c}{3} \log\left( \frac{\beta}{2\pi} 2\cosh\left(\frac{2\pi t}{\beta} \right)\right)~&,~~t \ll L/2 \, .\\ 
  \end{aligned}
  \right. 
  \label{eq:EE in TFD state}
\end{equation}
Here, time evolution is defined by evolving forwards in time in both copies,
\begin{equation}
  \ket{\Omega(t)} = e^{i(H_1 + H_2)t} \ket{\Omega}, 
\end{equation}
where $H_1,H_2$ are the Hamiltonians of the two copies\footnote{Evolving with $e^{i(H_1-H_2)t}$ leaves $\ket\Omega$ unchanged.}.

For $t \ll L/2$, \eqref{eq:EE in TFD state} is equal to the length of two bulk geodesics stretching between the two boundaries of the two-sided black hole and ending at the endpoints of the subintervals at both boundaries.
  For $t \gg L/2$, on the other hand, we have two bulk geodesics of the same kind as for the one sided case, i.e.~geodesics stretching between two endpoints of the same subinterval on one asymptotic boundary.
  As for the one sided-case, these geodesics wind $k$ times around the horizon and do not enter the black hole interior\footnote{Since there is a one-to-one map between geodesics in the BTZ geometry and geodesics on its covering space (pure AdS$_3$), we may equally well think of all of these geodesics as geodesics in pure AdS$_3$. Since the action of the bulk quotient does not change the local geometry, but only the global topology, the geodesic lengths in both cases agree.}.

Thus, eq.~\eqref{eq:EE in TFD state} shows that entwinement captures the growth of the wormhole for a time that depends linearly on $k \propto L$.
While the equality between geodesic lengths and \eqref{eq:EE in TFD state} is exact and holds for any $k \in \NN$, for entwinement $k$ is bounded from above by the condition $k \ll M \ll c$ that was necessary for the calculation to be well-defined.
Hence, the growth of entwinement with time is limited by $M$, whose value is itself limited by the central charge $c$.
However, for large enough $k$, it still continues for a much longer time than the growth of the entanglement entropy.
The upper limit on the time that entwinement can describe the wormhole growth imposed by $k \ll M \ll c$ is in agreement with the expectation that quantum corrections lead to a breakdown of the wormhole growth at late times \cite{Susskind:2014moa}.
Of course, stating this argument properly requires detailed knowledge of the microstates of the ensemble and cannot be done in the averaged description that we base our definition of entwinement on.

We also note that in the late time limit, the entanglement entropy in the covering theory saturates at the maximal value allowed by the positivity of mutual information\footnote{Since entwinement reduces to ordinary entanglement entropy in an enlarged Hilbert space which respects positivity of mutual information, we know that positivity of mutual information also holds for entwinement.} between the two components of the entangling interval (the entanglement entropy of one component is given by eq.~\eqref{eq:EE cylinder}).
Since the CFT in question is holographic and in the large central charge limit, the Ryu-Takayanagi formula applies and the phase transition at time $t = L/2$ is a first order transition.

In appendix \ref{sec:BTZ geodesics} we show that with a slight generalization of the calculation of \cite{Hartman:2013qma}, entwinement captures the length of goedesics that not only stretch between different asymptotic boundaries of the two-sided black hole but also wind a non-vanishing amount of times around the spatial circle\footnote{Note that these geodesics are different from the winding geodesics dual to entwinement introduced above. The winding geodesics described in \secref{sec:BTZ} are contained fully in one asymptotic region. In contrast, the geodesics derived in app.~\ref{sec:BTZ geodesics} pass through the wormhole from one asymptotic region to the other, while also winding $k$ times around the spatial circle.}.

Therefore, in summary we see that the definition of entwinement given in this section implies that the dual geodesics probe a region of spacetime which is much larger than the one probed by entanglement entropy.
The size of the bulk region probed by entwinement is limited by the value of the central charge $c$ in the boundary theory.
When the dual geodesics come close to the black hole horizon or singularity, the relation between geodesic lengths and entwinement breaks down.
This may be related to the fact that bulk quantum corrections could be necessary to describe the geometry in these regions.

\section{$S_n$ Orbifolds}
\label{sec:S_n Orbifolds} 
A further application of the techniques for defining entanglement entropy and entwinement of \secref{sec:entanglement entropy} and \ref{sec:entwinement} is possible in the D1/D5-brane system.
At a certain point in the moduli space the field theory description of this system is a $(T^4)^n/S_n$ symmetric product orbifold, where $n=n_1n_5$, $n_1$ is the number of D1-branes, $n_5$ the number of D5-branes and $c=6n_1n_5$ the central charge \cite{Strominger:1996sh,deBoer:1998kjm,Seiberg:1999xz,Larsen:1999uk}.
States of the $S_n$ orbifold are denoted by
\begin{equation}
  \ket{X^1,...,X^n}_h,
\end{equation}
where $X^i$ is the eigenvalue of the field operator $\hat X^i$ \footnote{When the difference between field operators and the corresponding eigenvalues in the states is not obvious, we use a hat on the field operators to distinguish them.} and the $h$ subscript indicates the boundary conditions $\hat X^i(\phi+2\pi)=\hat X^{h(i)}(\phi)$\footnote{We suppress he second index labeling the different fields in each of the $n$ copies (four free bosons and fermions in the $(T^4)^n/S_n$ case).}.
A $S_n$ generator $g$ acts by permuting $\hat X^i$ with $\hat X^{g(i)}$.
The boundary conditions change as
\begin{equation}
  \hat X^{g(i)}(\phi+2\pi) = \hat X^{gh(i)}(\phi)  ~\Leftrightarrow~ \hat X^{i}(\phi+2\pi) = \hat X^{ghg^{-1}(i)}(\phi).
\end{equation}
Therefore the action of $g$ is obtained as
\begin{equation}
  g\ket{X^1,...,X^n}_h = \ket{X^{g^{-1}(1)},...,X^{g^{-1}(n)}}_{ghg^{-1}},
  \label{eq:S_n transform state}
\end{equation}
where we have written $g$ in a slight abuse of notation for both the permutation operators acting on states in the Hilbert space as well as on numbers from 1 to $n$.
$S_n$ invariant states formed by applying the projection operator $P=\frac{1}{n!}\sum_{g \in S_n}g$ necessarily contain all states with boundary conditions in the same conjugacy class
\begin{equation}
  C(h) = \{\, ghg^{-1} ~|~ g \in S_n \,\},
\end{equation}
so twisted sectors are labeled not by a $S_n$ element $h$, but by a conjugacy class $C(h)$.
The group of $S_n$ elements commuting with a given $g$ is called the stabilizer subgroup $N(g)$.
The sum over $S_n$ splits up into a sum over $g \in C(h)$ that changes the boundary conditions and a sum over $N(g)$ that permutes some of the fields with fixed boundary conditions with each other.

Note that although the sum over elements of the conjugacy class $C(h)$ permutes some of the fields $\hat X^i$ with each other, it does not change the boundary conditions for the corresponding eigenvalues $X^i$.
The eigenvalues in $\ket{X^{g^{-1}(1)},...,X^{g^{-1}(n)}}_{ghg^{-1}}$ on the right hand side of \eqref{eq:S_n transform state} obey boundary conditions
\begin{equation}
  X^{g^{-1}(i)}(\phi+2\pi)=X^{g^{-1}ghg^{-1}(i)}(\phi)
  ~\Leftrightarrow~
  X^{i}(\phi+2\pi)=X^{h(i)}(\phi) 
\end{equation}
which are equivalent to those obeyed by the eigenvalues in $\ket{X^1,...,X^n}_h$ on the left hand side of \eqref{eq:S_n transform state}.
The situation in \cite{Balasubramanian:2018ajb} in which eigenvalues that were continuously connected by the boundary conditions became disconnected due to permutations does not occur in our case.

The orbifold Hilbert space $\cH_\text{orb}$ is a direct sum of Hilbert spaces consisting of elements with boundary conditions in the same conjugacy class.
Conjugacy classes for the $S_n$ group are specified by a partition of $n$ into integers,
\begin{equation}
  n = \sum_{N=1}^n N n_N.
\end{equation}
A generic state contains $n_N$ cycles of length $N$.
The corresponding stabilizer subgroups contain amongst other $S_n$ elements $\ZZ_N$ permutations of the fields inside all cycles of length $N$.
Thus, for a state with fixed boundary conditions $h \in S_n$ there exists an enlarged Hilbert space $\tilde\cH_g$ in which the fields inside the cycles of length $N$ are not $\ZZ_N$ symmetrized.
Embedding the density matrix into this enlarged Hilbert space and taking partial traces for tensor factors of $\tilde\cH_g$ gives a definition of entwinement.
From the results of \secref{sec:entwinement}, it is clear that this partial trace definition of entwinement can be reproduced from a definition employing a linear subspace of operators.

A particular example for a state of the $S_n$ orbifold is the one dual to the conical defect, which is the ground state of the twisted sector in which the fields are sewn together in $n/N$ cycles of length $N$ \cite{Balasubramanian:2005qu}.
In this case, $N(g)$ is given by the $S_n$ elements consisting of $\ZZ_N$ permutations inside the cycles of length $N$ and $S_{n/N}$ permutations of the cycles with each other \cite{Balasubramanian:2014sra}.
The enlarged non-$\ZZ_N$ invariant Hilbert space $\tilde\cH_g$, in which the $N(g)$ invariant states are embedded, is comprised of states invariant under the $S_{n/N}$.
This $S_{n/N}$ symmetry is the symmetry of states in the untwisted (vacuum) sector of a $S_{n/N}$ orbifold with central charge $\tilde c = 6n/N = c/N$.
Therefore, also for the $S_n$ orbifold entwinement in the state dual to the conical defect is given by the ordinary entanglement entropy in the vacuum state of the covering CFT with central charge $\tilde c=c/N$.

It would be very interesting to perform an analysis of entanglement entropy (see \cite{Asplund:2011cq,Giusto:2014aba,Giusto:2015dfa,Bombini:2019vuk} for some work in this direction) and particularly of entwinement for microstates of the BTZ black hole in the D1/D5-brane system. These microstates were studied for instance in the context of the fuzzball program \cite{Skenderis:2008qn,Mathur:2008nj}.
This would provide a strong test of the proposal of \secref{sec:BTZ}.

\section{Conclusions}
\label{sec:conclusions}
In this paper we have explored the entanglement structure of CFTs with discrete gauge symmetries, in particular for holographic CFTs dual to conical defects and the black holes in AdS$_3$.
We have shown that suitable generalizations of the concept of entanglement exist for observers that can measure only a subset of observables such as an algebra or a linear subspace of local operators.
These generalizations agree with naive calculations that work by embedding the density matrix into a larger factorizing Hilbert space containing unphysical states.
Futhermore, the corresponding entropies are equal to lengths of dual bulk geodesics.
In this way, we obtain an explicit construction of a in principle measureable quantity in the boundary field theory dual to the length of a non-minimal geodesic in bulk.
This enables us to obtain the full bulk geometry, including the entanglement shadow, from the field theory side.
Thus, entwinement offers a description of the features of the bulk geometry conjectured to be explained by complexity \cite{Susskind:2014moa,Susskind:2014rva}.

Entwinement, the quantity that measures entanglement between non-spatially organized degrees of freedom, has an analogue for systems of indistinguishable particles, as already observed in \cite{Balasubramanian:2018ajb}.
To see this, consider a system of $N$ indistinguishable qubits.
It is not possible to assign labels to the different qubits and to ask how a specific set of qubits, e.g.~qubit 1, 2, ... $k$, is entangled with the remainder.
However it is a valid question to ask how are $k$ particles entangled with the remaining $N-k$.
This is analogous to the question entwinement aims to answer.

Applying the same techniques as in \secref{sec:entwinement}, a generalized entanglement mesaure $E_k$ between $k$ qubits and the remaining $N-k$ in analogy to entwinement is defined as follows.
Consider performing measurements with all possible Hermitean operators that act on $k$ qubits simultaneously.
Since the qubits are indistinguishable, it is not possible to determine on which qubits these operators act on, but it is of course well-defined to state that they act on $k$ of them.
These measurements lead to probability distributions of states.
We identify the minimum of the entropy of the probability distributions with $E_k$.
We say that $k$ qubits are unentangled with the remaining $N-k$ if this entropy vanishes, i.e.~if we measure a unique pure state or equivalently if $E_k = 0$.

This definition, which for pure states is equivalent to the more general proposal of \cite{2003PhRvA..68c2308B,2004PhRvL..92j7902B,2005IJTP...44.2127B}, does not measure entanglement alone but also contains classical correlations from global mixed states and $\ZZ_N$ symmetry of the states.
In contrast, other proposals for quantifying entanglement in systems of indistinguishable particles (see amongst others \cite{2001PhRvA..64e4302L,2002AnPhy.299...88E,2001PhRvA..64d2310P,2014PhRvL.112o0501K,2003PhRvA..67b4301S,2003PhRvL..91i7902W,2004PhRvA..70a2109G,2009PhRvL.102q0503T}) often aim at differentiating correlations due to statistics and global mixed states from genuine quantum correlations (see e.g.~\cite{2002AnPhy.299...88E}).
Thus, the analogue of entwinement described above is not equivalent to such proposals.
Instead of measuring only the entropy of quantum correlations, it measures the entropy of all correlations observable from measurements acting on $k$ qubits simultaneously.
The motivation for this definition comes from ordinary entanglement entropy which -- although it measures a mixture of classical and quantum correlations -- has a simply geometric bulk dual for holographic CFTs.

\medskip

We also proposed a definition of entwinement for thermal states on a circle dual to the BTZ black hole.
Like entwinement for the state dual to the conical defect, the definition reduces to the entanglement entropy in a covering CFT on an $M$ times larger spatial circle with central charge $c/M$.
We take the covering theory to be also in a thermal state.
Then, entwinement calculates the length of a non-minimal geodesic in the bulk that probes the entanglement shadow around the event horizon of the black hole.
Furthermore, generalizing to the two-sided black hole case, the results of \cite{Hartman:2013qma} imply that entwinement also captures the growth of the wormhole with time in the dual bulk geometry.

While this definition yields a simple calculation procedure for entwinement, it rests on some assumptions about the microscopic nature of the degrees of freedom of the CFT state dual to the BTZ black hole.
Namely we assume that there exist some $M \in \NN$ such that the fields of the theory can be separated into $M$ subsets that are continuously connected by the boundary conditions, i.e.~we assume that the state is in an approriate twisted sector of an orbifold.
The black hole is dual to an ensemble of states, so this condition needs to hold only after averaging over the elements of the ensemble.
The larger the maximum possible value of $M$ is, the more of the bulk geometry is probed by the geodesics dual to entwinement.
It is an important next step to check this assumption in concrete examples.
An obvious candidate for this is the $S_n$ orbifold dual to the D1-D5 system, for which a large number of microstates have been identified as part of the fuzzball program (see \cite{Skenderis:2008qn,Mathur:2008nj} for a review).

However if these assumptions do hold, the results summarized in \secref{sec:BTZ} imply that entwinement probes a subset of the bulk spacetime that is much larger than that probed by the entanglement entropy of spatial subregions.
In particular, a striking implication is that entwinement is sensitive to the growth of the wormhole in the two-sided BTZ spacetime -- and thus sensitive to physics inside the horizon -- for a much longer time than entanglement entropy of spatial subregions.
Hence, the assertion that ``entanglement is not enough'' \cite{Susskind:2014moa} to describe the wormhole growth in two-sided black hole geometries is called into question, at least if one includes entanglement of non-spatially organized degrees of freedom.
In this way, entwinement provides an alternative to complexity proposed in \cite{Susskind:2014moa,Susskind:2014rva} that for AdS$_3$/CFT$_2$ could describe the same physics.
The advantage of entwinement is that unlike for complexity not only its definition in the boundary CFT is clear, but also its bulk dual.
In particular there are no free parameters in our proposal, whereas for complexity a clear picture for which reference state and gate set to choose to match holographic complexity definitions is still missing.

\medskip

Our work opens the door to  a large number of future directions.
First of all, it would be interesting to apply the notion of generalized entanglement to other examples such as higher dimensional asymptotically AdS geometries or more exotic bulk geometries such as the Bañados geometry \cite{Banados:1998gg}.
A further direction worth exploring is a possible generalization of entwinement to find a dual for geodesics not affixed to a constant time slice, in the spirit of the HRT proposal \cite{Hubeny:2007xt} (see also \cite{Hubeny:2014qwa} for the related concept of residual entropy). 

Moreover, for the state dual to the conical defect, the derived field theory dual to the lengths of non-minimal geodesics completes the field-theory description of the kinematic space of \cite{Czech:2015qta} comprised of boundary-anchored geodesics.
This kinematic space provides the basis for the reconstruction of geometric objects in the bulk from field-theory data using methods from integral geometry.
Our results thus pave the way to a new description of geometric objects such as curves, points or areas in the conical defect in terms of a field-theoretic quantity.
In particular, they allow for a derivation of a field-theory dual to holographic complexity for the conical defect, in generalization of the results of \cite{Abt:2017pmf,Abt:2018ywl} that for the vacuum state provide an expression of the volume in the complexity=volume proposal in terms of the entanglement entropy, using the kinematic space approach.

A further direction to explore is a possible application of entwinement to the explanation of quantum chaos from black holes in holography (see e.g.~\cite{Shenker:2013pqa,Shenker:2013yza,Leichenauer:2014nxa,Roberts:2014isa}).
The generalized entanglement of \cite{2005IJTP...44.2127B,2003PhRvA..68c2308B} equivalent to our entwinement definition for pure states was already proposed to be a good indicator of quantum chaos \cite{2006EL.....76..746W}.

\begin{acknowledgments}
We would like to thank Mari-Carmen Bañuls, Pascal Fries, René Meyer, Christian Northe and Ignacio A.~Reyes for discussions.
MG acknowledges financial support from the DFG through Würzburg-Dresden Cluster of Excellence on Complexity and Topology in Quantum Matter - ct.qmat (EXC 2147, project-id 39085490).
\end{acknowledgments}

\appendix 
\section{Winding geodesics in the two-sided black hole geometry}
\label{sec:BTZ geodesics}
We now derive that entwinement as defined in \secref{sec:BTZ} can as well compute the length of geodesics between different asymptotic boundaries of the wormhole in a two-sided BTZ black hole geometry with non-vanishing winding number around the spatial circle.
We first consider the field theory calculation in generalization of \cite{Hartman:2013qma} and then the dual geodesics.

\subsection{Field-theory calculation} 
Correlation functions in the TFD state $\ket{\Omega(t)}$ are calculated by an analytical continuation from Euclidean correlation functions on a background with periodic time direction \cite{Hartman:2013qma}.
Operators $\phi_1$ on the first boundary are inserted at $\tau=0$, while operators $\phi_2$ on the second boundary are inserted at $\tau=i\beta/2$ and analytically continued to $\tau=i\beta/2-2t$.

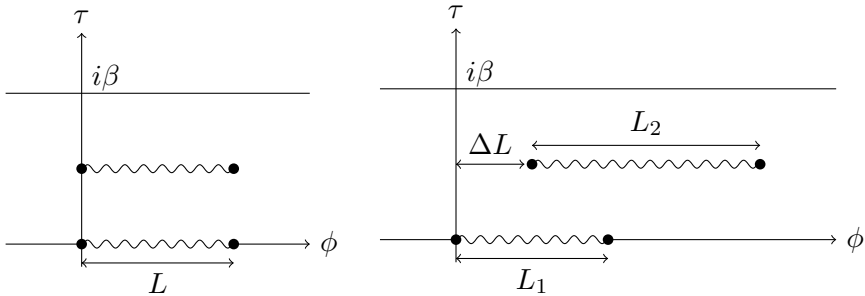
\begin{figure}
  \centering
  \begin{tabular}{cc}
    \begin{tikzpicture}
      \draw (-1,0) -- (0,0);
      \draw[->] (2,0) -- (3,0) node[right] {$\phi$};
      \draw[->] (0,-0.3) -- (0,2.8) node[above] {$\tau$};
      \draw (-1,2) -- (3,2);
      \draw (0,2.2) node[right] {$i\beta$};
      \draw[decorate,decoration={snake,amplitude=1.5,segment length=7}] (0,0) -- (2,0);
      \draw[decorate,decoration={snake,amplitude=1.5,segment length=7}] (0,1) -- (2,1);
      \fill (0,0) circle (0.07) (0,1) circle (0.07) (2,0) circle (0.07) (2,1) circle (0.07);
      \draw[<->] (0,-0.25) -- node[midway,below] {$L$} (2,-0.25);
    \end{tikzpicture} &
    \begin{tikzpicture}
      \draw (-1,0) -- (0,0);
      \draw[->] (2,0) -- (5,0) node[right] {$\phi$};
      \draw[->] (0,-0.3) -- (0,2.8) node[above] {$\tau$};
      \draw (-1,2) -- (5,2);
      \draw (0,2.2) node[right] {$i\beta$};
      \draw[decorate,decoration={snake,amplitude=1.5,segment length=7}] (0,0) -- (2,0);
      \draw[decorate,decoration={snake,amplitude=1.5,segment length=7}] (1,1) -- (4,1);
      \fill (0,0) circle (0.07) (2,0) circle (0.07) (1,1) circle (0.07) (4,1) circle (0.07);
      \draw[<->] (0,1) -- node[midway,above] {$\Delta L$} (0.9,1);
      \draw[<->] (1,1.25) -- node[midway,above] {$L_2$} (4,1.25);
      \draw[<->] (0,-0.25) -- node[midway,below] {$L_1$} (2,-0.25);
    \end{tikzpicture}
  \end{tabular}
  \caption{Twist operator insertions on the Euclidean cylinder for the computation of the entanglement entropy of an interval which consists of two disconnected components on the two asymptotic boundaries.
    LHS: computation of \cite{Hartman:2013qma}, both components of the entangling interval are of the same size and at the same position; RHS: position and size of both parts is arbitrary.}
  \label{fig:twist operators euclidean cylinder}
\end{figure}

Entwinement is given by the entanglement entropy in the covering space, which we will calculate from the Rényi entropy.
This entropy is given by a correlation function of twist operators,
\begin{equation}
  \Tr\rho_A^n = \langle \sigma_n(z_1,\bar z_1)
  \tilde\sigma_n(z_2,\bar z_2) \tilde\sigma_n(z_3,\bar z_3)
  \sigma_n(z_4,\bar z_4)
  \rangle.
\end{equation}
We want to calculate the most general entanglement entropy of an entangling interval consisting of two parts on both asymptotic boundaries.
Therefore, the entangling interval consist of a part with size $L_1$ on the first boundary and another part on the second boundary with size $L_2$  and relative position $\Delta L$ compared to the first part (see \figref{fig:twist operators euclidean cylinder}).
Thus,
\begin{equation}
  \begin{aligned}
    z_1 = \bar z_1 = 0 ~,~~ & z_3 = L_2 + \Delta L + 2t + i\beta/2~, ~~ \bar z_3 = L_2 + \Delta L - 2t - i\beta/2\\
    z_2 = \bar z_2 = L_1 ~,~~ & z_4 = - \bar z_4 = \Delta L + 2t + i\beta/2.
  \end{aligned}
\end{equation}

Mapping to the plane with $w=e^{2\pi z/\beta}$ and performing a conformal transformation maps the insertions to $0,x,1,\infty$, where $x$ is the conformally invariant cross-ratio.
For $L_1+L_2 \gg |L_1-L_2|$ and $L_1+L_2 \gg |L_1-L_2-2\Delta L|$, $x$ asymptotes to 0 for late times ($t \gg |L_1-L_2-2\Delta L|/4$) and to 1 for early times ($t \ll |L_1-L_2-2\Delta L|/4$).
Expanding the four-point function in conformal blocks and the blocks in a power series in $x$ resp. $1-x$, we see that the leading term is
\begin{equation}
  \langle \sigma_n(0,0) \tilde\sigma_n(x,\bar x) \tilde\sigma_n(1,1)
  \sigma_n(\infty,\infty) \rangle = \left\{
    \begin{aligned}
      x^{-2h}\bar x^{-2h}~,~~ t \gg |L_1-L_2-2\Delta L|/4\\
      (1-x)^{-2h}(1-\bar x)^{-2h}~,~~ t \ll |L_1-L_2-2\Delta L|/4
    \end{aligned}
  \right.
  \label{eq:leading term four-point function}
\end{equation}
where $h = \bar h = \frac{\tilde c}{24}(n-1/n)$ is the conformal weight of the twist operators.
Putting everything together, we obtain
\begin{equation}
  E^i_k = \left\{
    \begin{aligned}
      S_1(L_1) + S_1(L_2) ~,~~ t \gg |L_1-L_2-2\Delta L|/4\\
      S_2(\Delta L) + S_2(\Delta L+L_2-L_1) ~,~~ t \ll |L_1-L_2-2\Delta L|/4\\
    \end{aligned}
  \right.,
  \label{eq:EE in TFD state arbitrary intervals}
\end{equation}
where
\begin{equation}
  S_1(L) = \frac{\tilde c}{3} \log\left( \frac{\beta}{2\pi} 2\sinh\left(\frac{\pi L}{\beta} \right)\right) 
\end{equation}
is equal to the entanglement entropy for an interval of length $L$ in the thermal state and proportional to the length of a geodesic whose endpoints lie on the same asymptotic boundary whereas
\begin{equation}
  S_2(L) = \frac{\tilde c}{6} \log\left[ \left(\frac{\beta}{2\pi}\right)^2 \left(2\cosh\left(\frac{2\pi L}{\beta}\right) + 2\cosh\left(\frac{4\pi t}{\beta} \right)\right) \right]  
\end{equation}
is proportional to the length of a geodesic stretching between the two asymptotic boundaries and winding $\lfloor L/2\pi \rfloor$ times around the black hole, as we will see in the next subsection.
The behaviour changes if the conditions $L_1+L_2 \gg |L_1-L_2|, |L_1-L_2-2\Delta L|$ are not fulfilled, e.g.~for $L_1+L_2 \gg |L_1-L_2|$ but $L_1+L_2 \ll |L_1-L_2-2\Delta L|$ (large distance in the relative position) entwinement is given by $S_1(L_1) + S_2(L_2)$ for all times.

The observed growth of entanglement entropy in a thermofield double state on a line is in fact universal for every CFT.
The expansion in eq.~\eqref{eq:leading term four-point function} for early and late time is independent of the field content of the CFT.

\subsection{Dual geodesics}
To derive the length of geodesics dual to entwinement, we only need to calculate them in the covering geometry of the black hole geometry (i.e.~pure AdS$_3$), since the BTZ identification maps geodesics to geodesics.
Pure AdS$_3$ is given by the surface
\begin{equation}
  -Y_{-1}^2-Y_0^2+Y_1^2+Y_2^2 = -1
\end{equation}
in the embedding space $\RR^{2,2}$ with coordinates $Y_i$.
The time parameter which we have used in the previous calculations is defined w.r.t.~global AdS$_3$ coordinates given by
\begin{equation}
  \begin{aligned}
    Y_{-1} = \cosh x \cosh \rho ~~&~~ Y_0 = \sinh t \sinh \rho\\
    Y_2 = -\sinh x \cosh \rho ~~&~~ Y_1 = \cosh t \sinh \rho.
  \end{aligned}
\end{equation}
For simplicity, we will calculate the length of geodesics in Poincaré patch coordinates
\begin{equation}
  \begin{aligned}
    Y_{-1} = \frac{1}{2z}(1+(z^2-x_0^2+x_1^2)) ~~&~~ Y_0 = \frac{x_0}{z}\\
    Y_2 = \frac{1}{2z}(1-(z^2-x_0^2+x_1^2)) ~~&~~ Y_1 = \frac{x_1}{z}.
  \end{aligned}
\end{equation}
Finally, the winding number is defined w.r.t~the usual BTZ coordinates
\begin{equation}
  \begin{aligned}
    Y_{-1} = \frac{r}{r_+}\cosh(r_+\phi) ~~&~~ Y_0 = \sqrt{\frac{r^2}{r_+^2}-1}\sinh(r_+\tau)\\
    Y_2 = \sqrt{\frac{r^2}{r_+^2}-1}\cosh(r_+\tau) ~~&~~ Y_1 = \frac{r}{r_+}\sinh(r_+\phi).
  \end{aligned}
\end{equation}
The event horizon is located at $r_+ = \frac{2\pi}{\beta}$.
To simplify the notation, we set $r_+=1 \Leftrightarrow \beta=2\pi$ in the following.
The point $(t,x,\rho) = (t_b,L,\infty)$ maps to $(x_0,x_1,z) = (e^{-L}\sinh t_b,\pm e^{-L}\cosh t_b,0)$ \cite{Hartman:2013qma}, where the sign in the $x_1$ coordinate depends on which boundary of the wormhole the point is located at.
Therefore, we have to calculate the length of geodesics between the points
\begin{equation}
  \begin{aligned}
    P_1 = e^{-\Delta L}(\sinh t_b,-\cosh t_b,0) ~~&~~ P_3 = e^{-\Delta L-L_2}P_1\\
    P_2 = (\sinh t_b,\cosh t_b,0) ~~&~~ P_4 = e^{-L_1}P_2
  \end{aligned}
\end{equation}
in Poincaré patch coordinates.
Geodesics between two points $P^{(1)}=(x_0^{(1)},x_1^{(1)},0)$ and $P^{(2)}=(x_0^{(2)},x_1^{(2)},0)$ in this coordinate system are semi-circles that extend in the bulk up to a maximal radial coordinate $z^* = \frac{1}{2}\sqrt{(x_1^{(1)}-x_1^{(2)})^2-(x_0^{(1)}-x_0^{(2)})^2}$.
The length of a geodesic is only dependent on $z^*$ and the UV cutoff $\epsilon$,
\begin{equation}
  l_\gamma =
  2\int_\epsilon^{z^*}\frac{dz}{z}\frac{1}{\sqrt{1-(z/z^*)^2}} \approx 2\log\frac{2z^*}{\epsilon}.
\end{equation}
The dual bulk geodesics can either connect $P_1$ with $P_3$ and $P_2$ with $P_4$ or $P_1$ with $P_2$ and $P_3$ with $P_4$.
Given an appropriate cutoff and restoring the $\beta$ dependence, we obtain
\begin{equation}
  \begin{aligned}
    l_{P_1 \leftrightarrow P_3} &\sim \log\left(\sinh\left(\frac{\pi L_1}{\beta}\right)\right)\\
    l_{P_2 \leftrightarrow P_4} &\sim \log\left(\sinh\left(\frac{\pi L_2}{\beta}\right)\right)\\
    l_{P_1 \leftrightarrow P_2} &\sim \log\left(\cosh\left(\frac{2\pi \Delta L}{\beta}\right) + \cosh\left(\frac{4\pi t_b}{\beta}\right)\right)\\
    l_{P_3 \leftrightarrow P_4} &\sim \log\left(\cosh\left(\frac{2\pi (\Delta L+L_2-L_1)}{\beta}\right) + \cosh\left(\frac{4\pi t_b}{\beta}\right)\right).
  \end{aligned}
\end{equation}
The sum of these lengths $l_{P_1 \leftrightarrow P_3} + l_{P_2 \leftrightarrow P_4}$ and $l_{P_1 \leftrightarrow P_2} + l_{P_3 \leftrightarrow P_4}$ is equal - up to non-universal cutoff dependent factors - to the entwinement result \eqref{eq:EE in TFD state arbitrary intervals}.
At early times $|t| < t_b^*$, the $l_{P_1 \leftrightarrow P_2}$, $l_{P_3 \leftrightarrow P_4}$ geodesics form the Ryu-Takayanagi surface, while at late times $|t| > t_b^*$ the $l_{P_1 \leftrightarrow P_3}$, $l_{P_2 \leftrightarrow P_4}$ geodesics dominate.
The case $t_b^* = 0$ in which the RT surface is constant for all times is also possible, for example for large $\Delta L$ and $L_1 \approx L_2$.
To derive the winding number of the geodesics, we use the coordinate transformation to BTZ coordinates
\begin{equation}
  \cosh(r_+\phi) = \frac{1}{2}\left(\frac{1}{\sqrt{z^2-x_0^2+x_1^2}} + \sqrt{z^2-x_0^2+x_1^2}\right).
\end{equation}
Thus $\phi_{P_1} = -\Delta L$, $\phi_{P_2} = 0$, $\phi_{P_3} = -\Delta L-L_2$, $\phi_{P_4} = -L_1$. We use the difference $|\Delta\phi|$ in the $\phi$ coordinate at the endpoints of a geodesic to define its winding number $w = \lfloor |\Delta\phi|/2\pi \rfloor$.
At times $t_b < t_b^*$, the winding numbers are given by $w_{P_1 \leftrightarrow P_2} = \lfloor \Delta L/2\pi \rfloor$ and $w_{P_3 \leftrightarrow P_4} = \lfloor (\Delta L+L_2-L_1)/2\pi \rfloor$, while after the phase transition at $t_b > t_b^*$ the geodesics wind $\lfloor L_{1,2}/2\pi \rfloor$ times around the spatial circle.

\bibliographystyle{JHEP}
\bibliography{bibliography}

\end{document}